\documentclass[prl,reprint,twocolumn,showpacs,superscriptaddress]{revtex4-2}
\usepackage{float}
\usepackage{amsthm}
\usepackage{amssymb}
\usepackage{color}
\usepackage{booktabs}
\usepackage{makecell}
\usepackage{tabularx}
\usepackage{array}
\usepackage{graphicx}
\usepackage{subfigure}
\usepackage{stfloats}
\usepackage{multirow}
\usepackage{mathtools}
\usepackage{dcolumn}
\usepackage{bm}
\usepackage{algorithm2e}
\usepackage[T1]{fontenc}
\usepackage[colorlinks,
            linkcolor=blue,
            anchorcolor=blue,
            citecolor=blue,
            ]{hyperref}
\usepackage{xcolor}

\begin{document}

\title{Unified Framework for Calculating Convex Roof Resource Measures}

\author{Xuanran Zhu}
\thanks{These authors contributed equally to this work}
\affiliation{Department of Physics, The Hong Kong University of Science and Technology, Clear Water Bay, Kowloon, Hong Kong, China}

\author{Chao Zhang}
\thanks{These authors contributed equally to this work}
\affiliation{Department of Physics, The Hong Kong University of Science and Technology, Clear Water Bay, Kowloon, Hong Kong, China}

\author{Zheng An}
\affiliation{Department of Physics, The Hong Kong University of Science and Technology, Clear Water Bay, Kowloon, Hong Kong, China}

\author{Bei Zeng}
\email{zengb@ust.hk}
\affiliation{Department of Physics, The Hong Kong University of Science and Technology, Clear Water Bay, Kowloon, Hong Kong, China}

\date{\today}
\begin{abstract}
Quantum resource theories (QRTs) provide a comprehensive and practical framework for the analysis of diverse quantum phenomena. A fundamental task within QRTs is the quantification of resources inherent in a given quantum state. In this letter, we introduce a unified computational framework for a class of widely utilized quantum resource measures, derived from convex roof extensions. We establish that the computation of these convex roof resource measures can be reformulated as an optimization problem over a Stiefel manifold, which can be further unconstrained through polar projection. Compared to existing methods employing semi-definite programming (SDP), gradient-based techniques or seesaw strategy, our approach not only demonstrates superior computational efficiency but also maintains applicability across various scenarios within a streamlined workflow. We substantiate the efficacy of our method by applying it to several key quantum resources, including entanglement, coherence, and magic states. Moreover, our methodology can be readily extended to other convex roof quantities beyond the domain of resource theories, suggesting broad applicability in the realm of quantum information theory.
\end{abstract}

\maketitle

\emph{Introduction.}---
Many quantum phenomena have transitioned from being of purely theoretical interest to serving as invaluable \textit{resources} in quantum information processing tasks \cite{chitambar2019quantum}. These developments have prompted a rigorous investigation into the mathematical formulation of resource theories, with the aim of characterizing the quantum states and operations that can be harnessed for physical tasks. Originating with the study of quantum entanglement \cite{horodecki2009quantum}, quantum resource theories (QRTs) provide a comprehensive framework for analyzing athermality \cite{PhysRevLett.111.250404, gour2015resource}, asymmetry \cite{gour2008resource,PhysRevX.5.021001}, coherence \cite{aberg2006quantifying,PhysRevLett.113.140401,RevModPhys.89.041003}, nonclassicality \cite{PhysRevA.89.052302,PhysRevLett.119.230401}, magic states \cite{veitch2014resource,howard2017application,ahmadi2018quantification, PhysRevA.71.022316}, and other phenomena.

A QRT often denoted by the tuple $\mathcal{R}=(\mathcal{F},\mathcal{O})$, where the set $\mathcal{F}(\mathcal{H})$ denotes the free states in the given Hilbert space $\mathcal{H}$ and $\mathcal{O}$ is the set of free operations. The states not belonging to $\mathcal{F}$ are considered "resourceful", which are supposed to be quantified by a resource measure $R$. Two essential requirements for defining such measures are \textit{non-negativity} and \textit{monotonicity}: the former ensures that $R(\rho) \geq 0$ for any quantum state $\rho$ and the equality holds if and only if $\rho$ is a free state, while the latter guarantees that $R$ does not increase under free operations, i.e., $R(\Lambda(\rho)) \leq R(\rho)$ for any free operation $\Lambda \in \mathcal{O}$. Additional properties like strong monotonicity, additivity are also useful but not essential \cite{vidal2000entanglement,chitambar2019quantum}. 

\begin{figure}[H]
    \centering
    \includegraphics[width=\linewidth]{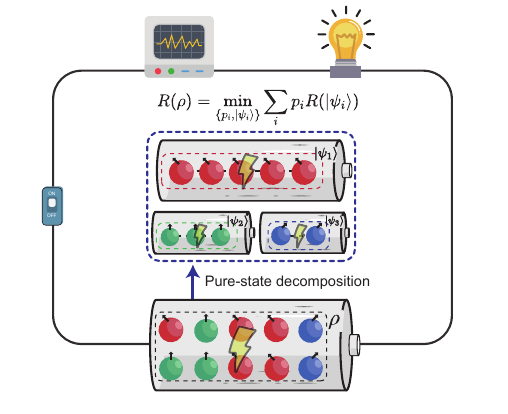}
    \caption{Convex roof resource measure $R(\rho)$ of a mixed state $\rho$ is calculated by minimizing the weighted sum of the resource measure $R(|\psi_i\rangle)$ over all possible pure-state decompositions.}
    \label{fig:pure_state_decomposition}
\end{figure}

One kind of widely used measures for quantifying the convex resource theories like entanglement, coherence, and magic states is constructed by the convex roof extensions \cite{PhysRevA.54.3824,uhlmann1998entropy,wei2003geometric,uhlmann2010roofs}, which begins with a measure of pure states and then extends to mixed states as the following form:
\begin{equation}
    R(\rho) = \min_{\{p_i,|\psi_i\rangle\}} \sum_i p_i R(|\psi_i\rangle),
\end{equation}
where the minimization is taken over all possible pure-state decompositions of the given mixed state $\rho$ satisfying $\rho = \sum_i p_i |\psi_i\rangle\langle\psi_i|$, as illustrated in Fig.~\ref{fig:pure_state_decomposition}.

In general, calculating resource measures is a challenging task and has been proven to be NP-hard in certain scenarios \cite{huang2014computing, horodecki2001entanglement}. Previous research has proposed various methods to compute the \textit{lower} or \textit{upper} bounds of these convex roof measures. The computation of lower bounds often relies on semi-definite programming (SDP), which relaxes pure-state decompositions into convex constraints \cite{toth2015evaluating, zhang2020numerical}. While SDP guarantees global minimum convergence for obtaining lower bounds, its computational complexity increases significantly with system size, rendering it impractical for relatively large-scale problems \cite{vandenberghe1996semidefinite}. And it may be incapable of handling certain scenarios like bound entanglement \cite{horodecki1998mixed}. Conversely, methods for computing upper bounds typically depend on gradient-based optimization \cite{audenaert2001variational, rothlisberger2009numerical}  or seesaw strategy \cite{streltsov2011simple}, which directly searches for the optimal pure-state decompositions. However, prior gradient-based studies require the manual derivation of the gradients of the objective function, significantly restricting their applicability to various scenarios. While the seesaw strategy encounters difficulties when applied to nearly pure multipartite states. Most critically, none of the aforementioned methods offer a unified framework for computing different convex roof resource measures.

In this letter, we propose a unified computational framework for various convex roof resource measures using the gradient descent method. We first demonstrate that the original problem of finding the optimal pure-state decomposition can be reformulated as an optimization problem over a Stiefel manifold. This problem can then be further unconstrained through the concept of trivialization \cite{lezcano2019trivializations}, wherein the free Euclidean space is mapped onto the desired Stiefel manifold via polar projection \cite{manton2002optimization}. Utilizing the automatic differentiation capabilities of PyTorch \cite{pytorch}, gradients can be obtained effortlessly without manual derivations, thereby enhancing the efficiency and applicability of our method within a straightforward workflow.

We validate the effectiveness of our approach by applying it to compute several key quantum resource measures, such as the entanglement of formation \cite{wootters2001entanglement}, the linear entropy of entanglement \cite{toth2015evaluating, buscemi2007linear}, the geometric measure of coherence \cite{PhysRevLett.115.020403, zhang2020numerical}, and the stabilizer entropy for quantifying the magic-state resource \cite{leone2022stabilizer, haug2023stabilizer, leone2024stabilizer}. Numerical results indicate that, in practice, our framework provides an effective way to compute accurate values of different convex roof resource measures, and at worst, provides \textit{upper} bounds. Moreover, our framework can be readily extended to other convex roof quantities, such as the constrained Holevo capacity in quantum communications \cite{shor2003capacities, shor2004equivalence} and quantum Fisher information (QFI) in the field of quantum metrology \cite{holevo2011probabilistic}. Consequently, it is anticipated that our method will find broad applications in the study of quantum information theories.

\emph{Method.}---
Given a quantum state with the eigenvalue decomposition $\rho=\sum_{i=1}^r \lambda_i |\lambda_i\rangle\langle\lambda_i|$, where $\langle \lambda_i|\lambda_j\rangle = \delta_{ij}$ and $r = \operatorname{rank}(\rho)$. One can verify that for any $n$-entry pure-state decomposition $\rho = \sum_{i=1}^n p_i |\psi_i\rangle\langle\psi_i|$, it can be represented by a series of auxiliary states $\{ |\tilde{\psi}_i\rangle \}_{i=1}^n$ satisfying $p_i= \langle \tilde{\psi}_i|\tilde{\psi}_i\rangle$ and $|\psi_i\rangle = \frac{1}{\sqrt{p_i}}|\tilde{\psi}_i\rangle$. These auxiliary states have the following form
\begin{equation}
    |\tilde{\psi}_i\rangle = \sum_{j=1}^r \sqrt{\lambda_j} X_{ij} |\lambda_j\rangle,
\end{equation}
where $X$ is a $n \times r$ complex matrix called \textit{Stiefel matrix}, satisfying $X^\dagger X = I$. The set of all $n \times r$ Stiefel matrices composes a special manifold called \textit{Stiefel manifold}, denoted by $\operatorname{St}(n,r)$. One direct result is that the optimization for a rank-$r$ state over the $n$-entry pure-state decomposition is equivalent to the optimization over the Stiefel manifold $\operatorname{St}(n,r)$, i.e.,
\begin{equation}
    \min_{\{p_i,|\psi_i\rangle\}} \sum_{i=1}^n p_i R(|\psi_i\rangle)= \min_{X \in \operatorname{St}(n,r)} \sum_{i=1}^n p_i(X)R(|\psi_i(X)\rangle).
\end{equation}

Fortunately, for the convex quantum resources where the free states compose a \textit{convex} set, the required decomposition number $n$ is bounded by the famous Carath\'{e}odory bound $d^2$ \cite{uhlmann2010roofs, streltsov2010linking}, where $d$ is the dimension of the given Hilbert space. Moreover, from the numerical experiments and previous analytical results \cite{lockhart2000optimal,wootters1998entanglement}, we find that in practice, the necessary decomposition number is often much smaller than the Carath\'{e}odory bound. Therefore, decomposition number $n$ can be a relatively small number during the optimization, which significantly reduces the computational complexity of the optimization.

Here, we utilize a parametrization strategy known as \textit{trivialization} \cite{lezcano2019trivializations} for solving the optimization problem over a manifold. The main idea is finding a surjective function $g$ mapping the free Euclidean space onto the desired manifolds. Here, we choose the polar projection for the trivialization of the Stiefel manifold \cite{manton2002optimization,chen2022tight}. It is defined as follows:
\begin{equation}
    g(A)=A(A^{\dagger}A)^{-\frac{1}{2}}: \mathbb{C}^{n\times r}\to\text{St}(n,r),
\end{equation}
where $2nr$ real parameters can be used to parametrize a $n \times r$ full-rank complex matrix $A$, while the dimension of the Stiefel manifold $\text{St}(n,r)$ is $2nr-r^2$. Numerically, it shows better performance compared to the previous parametrization strategies like the matrix exponential and Euler-Hurwitz angles \cite{audenaert2001variational, rothlisberger2009numerical}. We put the comparison of different trivialization mappings in the Supplementary Material \cite{supple}.

\begin{figure}[H]
    \centering
    \includegraphics[width=\linewidth]{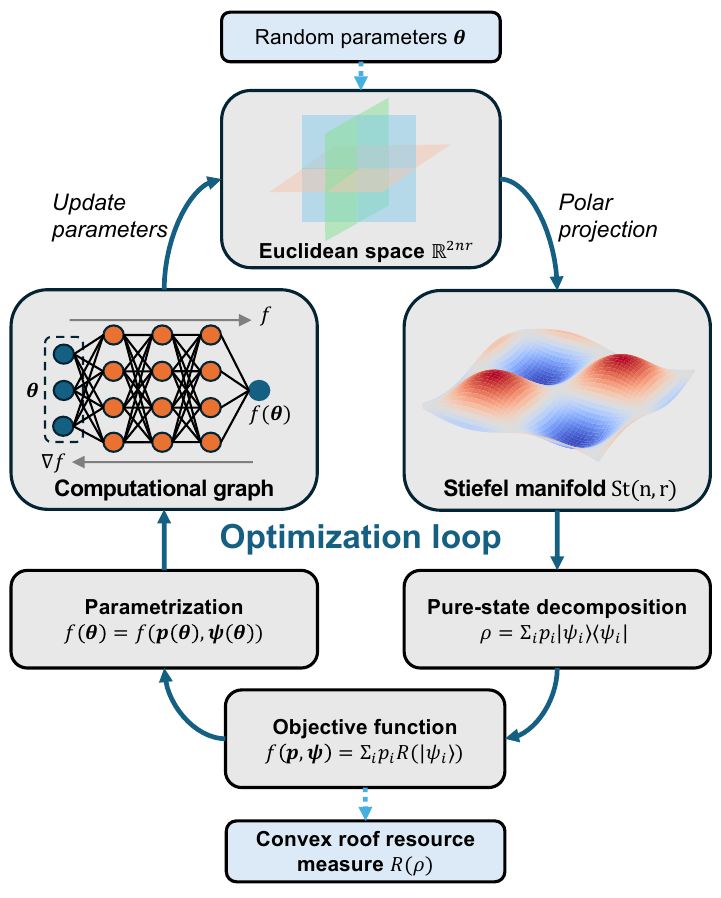}
    \caption{Flowchart of the unified framework for calculating convex roof resource measures. The optimization loop will stop when the objective function converges to the tolerance we set and then output the convex roof resource measure $R$ of the given state $\rho$.}
    \label{fig:flowchart}
\end{figure}

To minimize the objective function, we also adopt the gradient-based optimization method with the gradients obtained from gradient back-propagation provided in PyTorch deep learning framework \cite{pytorch}. Particularly, we choose the classical optimization method, L-BFGS implemented in SciPy \cite{virtanen2020scipy,10.1145/279232.279236} which shows better performance than the others. In Fig.~\ref{fig:flowchart}, we conclude the pivotal steps of the proposed framework in a flowchart. For different resource measures, we put the optimization details in the Supplementary Material \cite{supple}.

In the following, we will demonstrate the effectiveness of our method by applying it to compute several key quantum resource measures. For the consistency, we choose the same hyper-parameters unless otherwise stated, where the number of pure states for the decomposition is $n=2d$ ($d$ is the dimension of the given Hilbert), the repeat number for each optimization is $N=3$ for alleviating the local minimum issue and the tolerance of the optimization is set to be $10^{-14}$. These hyper-parameters are decided by numerical experiments but may not guarantee the optimality for certain cases. And all these results are produced on a standard laptop computer.

\emph{Resource of entanglement.}---
For quantifying entanglement, here, we introduce a famous entropic-based measure called entanglement of formation (EoF) \cite{wootters2001entanglement} and its linearized version -- linear entropy of entanglement \cite{toth2015evaluating, buscemi2007linear}. EoF is defined as the following form for a bipartite state $\rho \in \mathcal{H}_A\otimes \mathcal{H}_B$:
\begin{equation}
    E_f(\rho) = \min_{\{p_i,|\psi_i\rangle\}} \sum_i p_i E_f(|\psi_i\rangle),
\end{equation}
where $\{p_i,|\psi_i\rangle\}$ is the pure-state decomposition of $\rho$ and $E_f(|\psi\rangle)=S(\operatorname{Tr}_B[|\psi\rangle\langle\psi|])$ with the Von Neumann entropy $S(\rho)=-\operatorname{Tr}[\rho \ln \rho]$. As for the linear entropy of entanglement $E_l$, it can be obtained by simply replacing the von Neumann entropy $S(\rho)$ with the linear entropy $S_l(\rho)=1-\operatorname{Tr}[\rho^2]$. Both EoF and linear entropy of entanglement are qualified as entanglement measures satisfying the non-negativity and monotonicity under local operations and classical communication (LOCC) \cite{PhysRevA.67.012307,vidal2000entanglement}.

\begin{figure}[H]
    \centering
    \includegraphics[width=\linewidth]{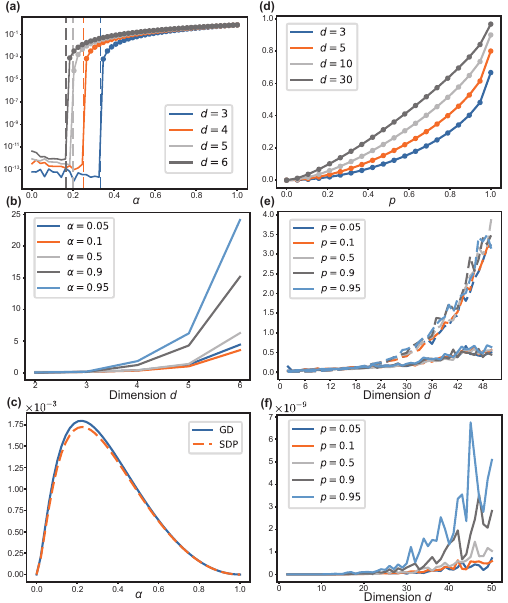}
    \caption{(a) Entanglement of formation \(E_f\) for \(d \otimes d\) Werner states \(\rho_W(\alpha)\). The vertical dashed lines indicate the exact boundaries between separable and entangled states.
    (b) Computational times (in seconds) for calculating \(E_f\) of different \(d \otimes d\) Werner states \(\rho_W(\alpha)\).
    (c) Linear entropy of entanglement \(E_l\) for \(3 \otimes 3\) bound entangled Horodecki states \(\rho_{\text{H}}(\alpha)\).
    (d) Geometric measure of coherence \(C_g\) for noisy coherent states indexed by the purity \(p\) with different dimensions $d$.
    (e) Computational times (in seconds) for calculating \(C_g\) of noisy coherent states indexed by \(p\) with different dimensions \(d\).
    (f) Numerical errors for computing \(C_g\) of noisy coherent states indexed by the purity \(p\) with different dimensions \(d\).
    In all figures, the solid curves represent numerical results obtained by gradient descent (GD), the dashed curves represent results from semi-definite programming (SDP), and the dots represent analytical results.}
    \label{fig:entanglement_coherence}
\end{figure}

In Fig.~\ref{fig:entanglement_coherence}(a), we compute the entanglement of formation (EoF) for the \(d \otimes d\) Werner states \(\rho_W(\alpha) = \frac{1}{d^2 - d\alpha}(I_{AB} - \alpha F_{AB})\), where \(I_{AB}\) is the identity operator and \(F_{AB} = \sum_{i,j=1}^d |i\rangle\langle j|_A \otimes |j\rangle \langle i|_B\) is the swap operator, acting on the Hilbert space $\mathcal{H}_A \otimes \mathcal{H}_B$. Analytical results, which are available in this case \cite{gao2008entanglement}, delineate the boundaries between separable and entangled states. Our method achieves values that are in close agreement with these analytical results. For each optimization, we use \(2nr \leq 4d^4\) parameters, implying that the computational time will increase polynomially with the local dimension $d$, consistent with the results in Fig.~\ref{fig:entanglement_coherence}(b).

Furthermore, we investigate the bound entanglement region in Fig.~\ref{fig:entanglement_coherence}(c) for the \(3 \otimes 3\) bound entangled states \(\rho_{\text{H}}(\alpha)\) introduced by Horodecki \cite{horodecki1997separability}. Our results demonstrate that our method can accurately determine the upper bounds of the linear entropy of entanglement, referred to the lower bounds obtained via semi-definite programming (SDP) \cite{toth2015evaluating}. In practice, we find that SDP approach can only handle the low-dimensional cases ($d_A, d_B \leq 3$) due to the demand of $O(d_A^4d_B^4)$ parameters. We put the relevant computational time comparisons and analysis in the Supplementary Material \cite{supple}.

\emph{Resource of coherence.}---
In a $d$-dimensional Hilbert space with its corresponding reference basis $\{|i\rangle\}$, a state is incoherent if and only if it is diagonal in the reference basis. Geometric measure of coherence is defined as follows
\begin{equation}
    C_g(\rho) = \min_{\{p_i,|\psi_i\rangle\}} \sum_i p_i C_g(|\psi_i\rangle),
\end{equation}
where $C_g(|\psi\rangle)=1-\max_{i}|\langle i|\psi\rangle|^2$ is the geometric measure of coherence for pure states. It fulfills the non-negativity and even strong monotonicity under incoherent operations \cite{PhysRevLett.115.020403}. We can also represent it equivalently as a distance-based measure \cite{streltsov2010linking}, i.e., $C_g(\rho)=1-\max_{\sigma\in \mathcal{F}}F(\rho,\sigma)$, where the fidelity $F(\rho,\sigma)=\|\sqrt{\rho}\sqrt{\sigma}\|_1^{2}$ and $\mathcal{F}$ is the set of incoherent states.

For the noisy coherent state $\rho=p|\psi^{+}\rangle\langle \psi^{+}|+(1-p)\frac{I}{d}$, where $|\psi^{+}\rangle=\frac{1}{\sqrt{d}}\sum_{i=1}^d|i\rangle$ is the maximally coherent state, its analytical solution of geometric measure of coherence is known as \cite{zhang2020numerical}
\begin{equation}
    C_g(\rho)=1-\frac{1}{d^2}[(d-1)\sqrt{1-p}+\sqrt{1+(d-1)p}]^2.
\end{equation}

In Fig.~\ref{fig:entanglement_coherence}(d), we calculate the geometric measure of coherence \(C_g\) for the noisy coherent states with different dimensions \(d\). The results show that our method achieves accurate values compared to the analytical results. In Fig.~\ref{fig:entanglement_coherence}(e), we plot the computational time for calculating \(C_g\) of the noisy coherent states with different dimensions \(d\), using gradient descent (GD) and semi-definite programming (SDP) \cite{zhang2020numerical}. The results indicate that the computational time for SDP increases significantly more than that for GD and the relevant time-scaling analysis is available in the Supplementary Material \cite{supple}. We also investigate the numerical errors during the computation, as shown in Fig.~\ref{fig:entanglement_coherence}(f). The results suggest that although the numerical errors increase more significantly for noisy coherent states with larger purity \(p\) as the dimension \(d\) increases, all numerical errors remain below \(10^{-8}\) for relatively large dimensions.

\emph{Resource of magic-state.}---
The inception of magic-state resource theory can be traced back to the seminal work of Bravyi and Kitaev \cite{PhysRevA.71.022316}, and then formalized in \cite{veitch2014resource}. In this scenario, stabilizer operations are considered as free operations, and the stabilizer states are considered as free states while non-stabilizer states (or magic states) are regarded as resourceful.

Stabilizer entropies (SEs) characterized by a $\alpha$-R\'{e}nyi index have recently been proposed to probe nonstabilizerness in multiqubit states \cite{leone2022stabilizer}. They have garnered attention due to their analytical properties \cite{PhysRevA.106.042426,PhysRevB.107.035148}, numerical computability \cite{PhysRevLett.131.180401} and experimental measurability \cite{oliviero2022measuring}.

The $\alpha$-R\'enyi stabilizer entropy of a pure state $|\psi\rangle$ on $n$ qubits is defined as
\begin{equation}
    M_{\alpha}(|\psi\rangle)=\frac{1}{1-\alpha}\log P_{\alpha}(|\psi\rangle),
\end{equation}
where $P_{\alpha}(|\psi\rangle)=\frac{1}{d_n}\sum_{P\in \mathbb{P}_n}|\langle\psi|P|\psi\rangle|^{2\alpha}$ is referred to as stabilizer purity, $d_n=2^n$ is the dimension of $n$-qubit Hilbert space, and $\mathbb{P}_n$ denotes the set of $n$-qubit Pauli operators. However, there is a counterexample to the monotonicity of stabilizer entropies with $\alpha < 2$ under stabilizer protocols \cite{haug2023stabilizer}. Recently, it was proved that stabilizer entropies with $\alpha \geq 2$ are monotones for magic-state resource theory when restricted to pure states \cite{leone2024stabilizer}. And their linearized versions -- linear stabilizer entropies -- serve as strong monotones, which have the form $M_{\alpha}^{\text{lin}}(|\psi\rangle)=1-P_{\alpha}(|\psi\rangle)$.

Similarly, they can be extended to the mixed states by the convex roof extension, i.e., $M_{\alpha}(\rho)=\frac{1}{1-\alpha}\log P_{\alpha}(\rho)$ and $M_{\alpha}^{\text{lin}}(\rho)=1-P_{\alpha}(\rho)$, where
\begin{equation}
    P_{\alpha}(\rho)= \max_{\{p_i,|\psi_i\rangle \}} \sum_i p_i P_{\alpha}(|\psi_i\rangle).
\end{equation}

\begin{figure}[H]
    \centering
    \includegraphics[width=\linewidth]{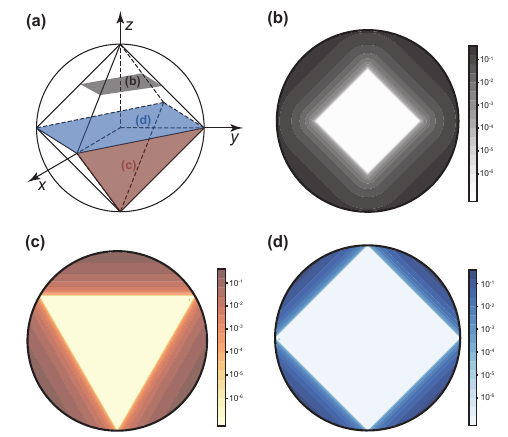}
    \caption{The octahedron inscribed in the Bloch sphere represents the region of stabilizer states. The colors in (b)(c)(d) indicate the position of those cross sections in the Bloch sphere, as shown in (a). Through the heatmap of the linear stabilizer entropy $M_{\alpha}^{\text{lin}}(\rho)$ with $\alpha=2$, the boundary between the stabilizer states and magic states is clearly shown.}
    \label{fig:magic_qubit}
\end{figure}

For testing the effectiveness of our method, we calculate the linear stabilizer entropy of the qubit states. As proven in \cite{PhysRevA.71.022316}, the stabilizer states in the qubit space compose an octahedron inscribed in the Bloch sphere, where the vertices are the pure stabilizer states. We plot three cross sections of the Bloch sphere in Fig.~\ref{fig:magic_qubit}, where the values of the linear stabilizer entropy with $\alpha=2$ are presented in the heatmaps. The results show clear boundaries between the stabilizer states and magic states, which is consistent with the analytical results.

\emph{Beyond resource theories.}---
The framework we proposed can also be extended to other forms of convex roof quantities beyond the resource theories. For example, in quantum communication theory, the Holevo capacity $\chi$ is an important quantity for characterizing the classical capacity of a quantum channel $\mathcal{N}$ \cite{holevo1998capacity,schumacher1997sending}, defined as $\chi(\mathcal{N})=\max_{\rho} \chi_{\mathcal{N}}(\rho)$, where the constrained Holevo capacity
\begin{equation}
    \chi_{\mathcal{N}}(\rho)=\max_{\{p_i,|\psi_i\rangle\}}S(\mathcal{N}(\rho))-\sum_i p_i S(\mathcal{N}(|\psi_i\rangle\langle\psi_i|)).
\end{equation}
Here, $\{p_i,|\psi_i\rangle\}$ is the pure-state decomposition of the input state $\rho$, and $S(\rho)=-\operatorname{Tr}[\rho\ln\rho]$ is the von Neumann entropy. It is clear that the constrained Holevo capacity can be calculated by the same workflow as the resource measures.

Another example is the quantum Fisher information (QFI) \cite{holevo2011probabilistic}, which places the fundamental limit to the accuracy of estimating an unknown parameter, playing an important role in quantum metrology \cite{giovannetti2004quantum,giovannetti2006quantum}. Taking the task of estimating the time parameter $t$ in the quantum evolution $U=e^{-iHt}$ as an instance, it holds the quantum Cramer-Rao bound $\Delta t F(\rho, H) \geq 1$, where $\Delta t$ characterizes the accuracy of estimating $t$ by any possible measurement made on the quantum state $U \rho U^{\dagger}$. It was firstly conjectured that QFI can also be expressed by convex roof extension \cite{toth2013extremal} and then proved in \cite{yu2013quantum}, i.e.,
\begin{equation}
    F(\rho, H)= 4 \min_{\{p_i,|\psi_i\rangle\}} \sum_i p_i (\Delta H)^2_{\psi_i},
\end{equation}
where $(\Delta H)^2_{\psi}=\langle \psi | H^2|\psi\rangle - \langle \psi |H | \psi \rangle ^2$ is the variance of $H$ by measuring a pure state $|\psi\rangle$. QFI can be applied to other scenarios via the generalization of variance such as the standard symmetrized variance introduced in \cite{zhao2022standard}. Our framework also provides a feasible way to calculate that kind of convex roof quantities.

\emph{Conclusion.}---
We have introduced a unified framework for calculating convex roof resource measures and demonstrated its effectiveness by applying it to several key quantum resources, including entanglement, coherence, and magic states. Our results underscore the efficiency and accuracy of our method in computing these measures. Even when our method encounters local minima, it still provides upper bounds, though such occurrences are rare in practice. Since semi-definite programming (SDP) offers lower bounds and our method provides upper bounds, combining these approaches is advisable for more precise computations of specific measures, as illustrated in Fig.~\ref{fig:entanglement_coherence}(c). However, the substantial computational resources required by the SDP approach may render it impractical for relatively large systems. In contrast, our gradient-based method, which leverages advanced non-convex optimization techniques, can effectively manage larger systems and yield valuable insights. Moreover, the versatility of our framework extends beyond resource theories, applying to other convex roof quantities such as constrained Holevo capacity and quantum Fisher information. This indicates its broad applicability in the study of quantum information theories. All results presented are reproducible using the code available in our public repository \cite{zhu_2024_12540301}. For more discussions and details, please refer to the Supplementary Material \cite{supple}.

\begin{acknowledgments}
We gratefully acknowledge the discussion with Jiahui Wu and the assistance of ChatGPT in facilitating the writing process. X-R. Zhu, C. Zhang, Z. An and B. Zeng are supported by GRF grant No. 16305121.
\end{acknowledgments}

\bibliography{reference}

\begin{thebibliography}{69}%
\makeatletter
\providecommand \@ifxundefined [1]{%
 \@ifx{#1\undefined}
}%
\providecommand \@ifnum [1]{%
 \ifnum #1\expandafter \@firstoftwo
 \else \expandafter \@secondoftwo
 \fi
}%
\providecommand \@ifx [1]{%
 \ifx #1\expandafter \@firstoftwo
 \else \expandafter \@secondoftwo
 \fi
}%
\providecommand \natexlab [1]{#1}%
\providecommand \enquote  [1]{``#1''}%
\providecommand \bibnamefont  [1]{#1}%
\providecommand \bibfnamefont [1]{#1}%
\providecommand \citenamefont [1]{#1}%
\providecommand \href@noop [0]{\@secondoftwo}%
\providecommand \href [0]{\begingroup \@sanitize@url \@href}%
\providecommand \@href[1]{\@@startlink{#1}\@@href}%
\providecommand \@@href[1]{\endgroup#1\@@endlink}%
\providecommand \@sanitize@url [0]{\catcode `\\12\catcode `\$12\catcode `\&12\catcode `\#12\catcode `\^12\catcode `\_12\catcode `\%12\relax}%
\providecommand \@@startlink[1]{}%
\providecommand \@@endlink[0]{}%
\providecommand \url  [0]{\begingroup\@sanitize@url \@url }%
\providecommand \@url [1]{\endgroup\@href {#1}{\urlprefix }}%
\providecommand \urlprefix  [0]{URL }%
\providecommand \Eprint [0]{\href }%
\providecommand \doibase [0]{https://doi.org/}%
\providecommand \selectlanguage [0]{\@gobble}%
\providecommand \bibinfo  [0]{\@secondoftwo}%
\providecommand \bibfield  [0]{\@secondoftwo}%
\providecommand \translation [1]{[#1]}%
\providecommand \BibitemOpen [0]{}%
\providecommand \bibitemStop [0]{}%
\providecommand \bibitemNoStop [0]{.\EOS\space}%
\providecommand \EOS [0]{\spacefactor3000\relax}%
\providecommand \BibitemShut  [1]{\csname bibitem#1\endcsname}%
\let\auto@bib@innerbib\@empty
\bibitem [{\citenamefont {Chitambar}\ and\ \citenamefont {Gour}(2019)}]{chitambar2019quantum}%
  \BibitemOpen
  \bibfield  {author} {\bibinfo {author} {\bibfnamefont {E.}~\bibnamefont {Chitambar}}\ and\ \bibinfo {author} {\bibfnamefont {G.}~\bibnamefont {Gour}},\ }\bibfield  {title} {\bibinfo {title} {Quantum resource theories},\ }\href {https://doi.org/10.1103/RevModPhys.91.025001} {\bibfield  {journal} {\bibinfo  {journal} {Rev. Mod. Phys.}\ }\textbf {\bibinfo {volume} {91}},\ \bibinfo {pages} {025001} (\bibinfo {year} {2019})}\BibitemShut {NoStop}%
\bibitem [{\citenamefont {Horodecki}\ \emph {et~al.}(2009)\citenamefont {Horodecki}, \citenamefont {Horodecki}, \citenamefont {Horodecki},\ and\ \citenamefont {Horodecki}}]{horodecki2009quantum}%
  \BibitemOpen
  \bibfield  {author} {\bibinfo {author} {\bibfnamefont {R.}~\bibnamefont {Horodecki}}, \bibinfo {author} {\bibfnamefont {P.}~\bibnamefont {Horodecki}}, \bibinfo {author} {\bibfnamefont {M.}~\bibnamefont {Horodecki}},\ and\ \bibinfo {author} {\bibfnamefont {K.}~\bibnamefont {Horodecki}},\ }\bibfield  {title} {\bibinfo {title} {Quantum entanglement},\ }\href {https://doi.org/10.1103/RevModPhys.81.865} {\bibfield  {journal} {\bibinfo  {journal} {Rev. Mod. Phys.}\ }\textbf {\bibinfo {volume} {81}},\ \bibinfo {pages} {865} (\bibinfo {year} {2009})}\BibitemShut {NoStop}%
\bibitem [{\citenamefont {Brand\~ao}\ \emph {et~al.}(2013)\citenamefont {Brand\~ao}, \citenamefont {Horodecki}, \citenamefont {Oppenheim}, \citenamefont {Renes},\ and\ \citenamefont {Spekkens}}]{PhysRevLett.111.250404}%
  \BibitemOpen
  \bibfield  {author} {\bibinfo {author} {\bibfnamefont {F.~G. S.~L.}\ \bibnamefont {Brand\~ao}}, \bibinfo {author} {\bibfnamefont {M.}~\bibnamefont {Horodecki}}, \bibinfo {author} {\bibfnamefont {J.}~\bibnamefont {Oppenheim}}, \bibinfo {author} {\bibfnamefont {J.~M.}\ \bibnamefont {Renes}},\ and\ \bibinfo {author} {\bibfnamefont {R.~W.}\ \bibnamefont {Spekkens}},\ }\bibfield  {title} {\bibinfo {title} {Resource theory of quantum states out of thermal equilibrium},\ }\href {https://doi.org/10.1103/PhysRevLett.111.250404} {\bibfield  {journal} {\bibinfo  {journal} {Phys. Rev. Lett.}\ }\textbf {\bibinfo {volume} {111}},\ \bibinfo {pages} {250404} (\bibinfo {year} {2013})}\BibitemShut {NoStop}%
\bibitem [{\citenamefont {Gour}\ \emph {et~al.}(2015)\citenamefont {Gour}, \citenamefont {Müller}, \citenamefont {Narasimhachar}, \citenamefont {Spekkens},\ and\ \citenamefont {{Yunger Halpern}}}]{gour2015resource}%
  \BibitemOpen
  \bibfield  {author} {\bibinfo {author} {\bibfnamefont {G.}~\bibnamefont {Gour}}, \bibinfo {author} {\bibfnamefont {M.~P.}\ \bibnamefont {Müller}}, \bibinfo {author} {\bibfnamefont {V.}~\bibnamefont {Narasimhachar}}, \bibinfo {author} {\bibfnamefont {R.~W.}\ \bibnamefont {Spekkens}},\ and\ \bibinfo {author} {\bibfnamefont {N.}~\bibnamefont {{Yunger Halpern}}},\ }\bibfield  {title} {\bibinfo {title} {The resource theory of informational nonequilibrium in thermodynamics},\ }\href {https://doi.org/https://doi.org/10.1016/j.physrep.2015.04.003} {\bibfield  {journal} {\bibinfo  {journal} {Physics Reports}\ }\textbf {\bibinfo {volume} {583}},\ \bibinfo {pages} {1} (\bibinfo {year} {2015})}\BibitemShut {NoStop}%
\bibitem [{\citenamefont {Gour}\ and\ \citenamefont {Spekkens}(2008)}]{gour2008resource}%
  \BibitemOpen
  \bibfield  {author} {\bibinfo {author} {\bibfnamefont {G.}~\bibnamefont {Gour}}\ and\ \bibinfo {author} {\bibfnamefont {R.~W.}\ \bibnamefont {Spekkens}},\ }\bibfield  {title} {\bibinfo {title} {The resource theory of quantum reference frames: manipulations and monotones},\ }\href {https://doi.org/10.1088/1367-2630/10/3/033023} {\bibfield  {journal} {\bibinfo  {journal} {New Journal of Physics}\ }\textbf {\bibinfo {volume} {10}},\ \bibinfo {pages} {033023} (\bibinfo {year} {2008})}\BibitemShut {NoStop}%
\bibitem [{\citenamefont {Lostaglio}\ \emph {et~al.}(2015)\citenamefont {Lostaglio}, \citenamefont {Korzekwa}, \citenamefont {Jennings},\ and\ \citenamefont {Rudolph}}]{PhysRevX.5.021001}%
  \BibitemOpen
  \bibfield  {author} {\bibinfo {author} {\bibfnamefont {M.}~\bibnamefont {Lostaglio}}, \bibinfo {author} {\bibfnamefont {K.}~\bibnamefont {Korzekwa}}, \bibinfo {author} {\bibfnamefont {D.}~\bibnamefont {Jennings}},\ and\ \bibinfo {author} {\bibfnamefont {T.}~\bibnamefont {Rudolph}},\ }\bibfield  {title} {\bibinfo {title} {Quantum coherence, time-translation symmetry, and thermodynamics},\ }\href {https://doi.org/10.1103/PhysRevX.5.021001} {\bibfield  {journal} {\bibinfo  {journal} {Phys. Rev. X}\ }\textbf {\bibinfo {volume} {5}},\ \bibinfo {pages} {021001} (\bibinfo {year} {2015})}\BibitemShut {NoStop}%
\bibitem [{\citenamefont {Aberg}(2006)}]{aberg2006quantifying}%
  \BibitemOpen
  \bibfield  {author} {\bibinfo {author} {\bibfnamefont {J.}~\bibnamefont {Aberg}},\ }\bibfield  {title} {\bibinfo {title} {Quantifying superposition},\ }\href@noop {} {\bibfield  {journal} {\bibinfo  {journal} {arXiv preprint quant-ph/0612146}\ } (\bibinfo {year} {2006})}\BibitemShut {NoStop}%
\bibitem [{\citenamefont {Baumgratz}\ \emph {et~al.}(2014)\citenamefont {Baumgratz}, \citenamefont {Cramer},\ and\ \citenamefont {Plenio}}]{PhysRevLett.113.140401}%
  \BibitemOpen
  \bibfield  {author} {\bibinfo {author} {\bibfnamefont {T.}~\bibnamefont {Baumgratz}}, \bibinfo {author} {\bibfnamefont {M.}~\bibnamefont {Cramer}},\ and\ \bibinfo {author} {\bibfnamefont {M.~B.}\ \bibnamefont {Plenio}},\ }\bibfield  {title} {\bibinfo {title} {Quantifying coherence},\ }\href {https://doi.org/10.1103/PhysRevLett.113.140401} {\bibfield  {journal} {\bibinfo  {journal} {Phys. Rev. Lett.}\ }\textbf {\bibinfo {volume} {113}},\ \bibinfo {pages} {140401} (\bibinfo {year} {2014})}\BibitemShut {NoStop}%
\bibitem [{\citenamefont {Streltsov}\ \emph {et~al.}(2017)\citenamefont {Streltsov}, \citenamefont {Adesso},\ and\ \citenamefont {Plenio}}]{RevModPhys.89.041003}%
  \BibitemOpen
  \bibfield  {author} {\bibinfo {author} {\bibfnamefont {A.}~\bibnamefont {Streltsov}}, \bibinfo {author} {\bibfnamefont {G.}~\bibnamefont {Adesso}},\ and\ \bibinfo {author} {\bibfnamefont {M.~B.}\ \bibnamefont {Plenio}},\ }\bibfield  {title} {\bibinfo {title} {Colloquium: Quantum coherence as a resource},\ }\href {https://doi.org/10.1103/RevModPhys.89.041003} {\bibfield  {journal} {\bibinfo  {journal} {Rev. Mod. Phys.}\ }\textbf {\bibinfo {volume} {89}},\ \bibinfo {pages} {041003} (\bibinfo {year} {2017})}\BibitemShut {NoStop}%
\bibitem [{\citenamefont {Vogel}\ and\ \citenamefont {Sperling}(2014)}]{PhysRevA.89.052302}%
  \BibitemOpen
  \bibfield  {author} {\bibinfo {author} {\bibfnamefont {W.}~\bibnamefont {Vogel}}\ and\ \bibinfo {author} {\bibfnamefont {J.}~\bibnamefont {Sperling}},\ }\bibfield  {title} {\bibinfo {title} {Unified quantification of nonclassicality and entanglement},\ }\href {https://doi.org/10.1103/PhysRevA.89.052302} {\bibfield  {journal} {\bibinfo  {journal} {Phys. Rev. A}\ }\textbf {\bibinfo {volume} {89}},\ \bibinfo {pages} {052302} (\bibinfo {year} {2014})}\BibitemShut {NoStop}%
\bibitem [{\citenamefont {Theurer}\ \emph {et~al.}(2017)\citenamefont {Theurer}, \citenamefont {Killoran}, \citenamefont {Egloff},\ and\ \citenamefont {Plenio}}]{PhysRevLett.119.230401}%
  \BibitemOpen
  \bibfield  {author} {\bibinfo {author} {\bibfnamefont {T.}~\bibnamefont {Theurer}}, \bibinfo {author} {\bibfnamefont {N.}~\bibnamefont {Killoran}}, \bibinfo {author} {\bibfnamefont {D.}~\bibnamefont {Egloff}},\ and\ \bibinfo {author} {\bibfnamefont {M.~B.}\ \bibnamefont {Plenio}},\ }\bibfield  {title} {\bibinfo {title} {Resource theory of superposition},\ }\href {https://doi.org/10.1103/PhysRevLett.119.230401} {\bibfield  {journal} {\bibinfo  {journal} {Phys. Rev. Lett.}\ }\textbf {\bibinfo {volume} {119}},\ \bibinfo {pages} {230401} (\bibinfo {year} {2017})}\BibitemShut {NoStop}%
\bibitem [{\citenamefont {Veitch}\ \emph {et~al.}(2014)\citenamefont {Veitch}, \citenamefont {Mousavian}, \citenamefont {Gottesman},\ and\ \citenamefont {Emerson}}]{veitch2014resource}%
  \BibitemOpen
  \bibfield  {author} {\bibinfo {author} {\bibfnamefont {V.}~\bibnamefont {Veitch}}, \bibinfo {author} {\bibfnamefont {S.~A.~H.}\ \bibnamefont {Mousavian}}, \bibinfo {author} {\bibfnamefont {D.}~\bibnamefont {Gottesman}},\ and\ \bibinfo {author} {\bibfnamefont {J.}~\bibnamefont {Emerson}},\ }\bibfield  {title} {\bibinfo {title} {The resource theory of stabilizer quantum computation},\ }\href {https://doi.org/10.1088/1367-2630/16/1/013009} {\bibfield  {journal} {\bibinfo  {journal} {New Journal of Physics}\ }\textbf {\bibinfo {volume} {16}},\ \bibinfo {pages} {013009} (\bibinfo {year} {2014})}\BibitemShut {NoStop}%
\bibitem [{\citenamefont {Howard}\ and\ \citenamefont {Campbell}(2017)}]{howard2017application}%
  \BibitemOpen
  \bibfield  {author} {\bibinfo {author} {\bibfnamefont {M.}~\bibnamefont {Howard}}\ and\ \bibinfo {author} {\bibfnamefont {E.}~\bibnamefont {Campbell}},\ }\bibfield  {title} {\bibinfo {title} {Application of a resource theory for magic states to fault-tolerant quantum computing},\ }\href {https://doi.org/10.1103/PhysRevLett.118.090501} {\bibfield  {journal} {\bibinfo  {journal} {Phys. Rev. Lett.}\ }\textbf {\bibinfo {volume} {118}},\ \bibinfo {pages} {090501} (\bibinfo {year} {2017})}\BibitemShut {NoStop}%
\bibitem [{\citenamefont {Ahmadi}\ \emph {et~al.}(2018)\citenamefont {Ahmadi}, \citenamefont {Dang}, \citenamefont {Gour},\ and\ \citenamefont {Sanders}}]{ahmadi2018quantification}%
  \BibitemOpen
  \bibfield  {author} {\bibinfo {author} {\bibfnamefont {M.}~\bibnamefont {Ahmadi}}, \bibinfo {author} {\bibfnamefont {H.~B.}\ \bibnamefont {Dang}}, \bibinfo {author} {\bibfnamefont {G.}~\bibnamefont {Gour}},\ and\ \bibinfo {author} {\bibfnamefont {B.~C.}\ \bibnamefont {Sanders}},\ }\bibfield  {title} {\bibinfo {title} {Quantification and manipulation of magic states},\ }\href {https://doi.org/10.1103/PhysRevA.97.062332} {\bibfield  {journal} {\bibinfo  {journal} {Phys. Rev. A}\ }\textbf {\bibinfo {volume} {97}},\ \bibinfo {pages} {062332} (\bibinfo {year} {2018})}\BibitemShut {NoStop}%
\bibitem [{\citenamefont {Bravyi}\ and\ \citenamefont {Kitaev}(2005)}]{PhysRevA.71.022316}%
  \BibitemOpen
  \bibfield  {author} {\bibinfo {author} {\bibfnamefont {S.}~\bibnamefont {Bravyi}}\ and\ \bibinfo {author} {\bibfnamefont {A.}~\bibnamefont {Kitaev}},\ }\bibfield  {title} {\bibinfo {title} {Universal quantum computation with ideal clifford gates and noisy ancillas},\ }\href {https://doi.org/10.1103/PhysRevA.71.022316} {\bibfield  {journal} {\bibinfo  {journal} {Phys. Rev. A}\ }\textbf {\bibinfo {volume} {71}},\ \bibinfo {pages} {022316} (\bibinfo {year} {2005})}\BibitemShut {NoStop}%
\bibitem [{\citenamefont {Vidal}(2000)}]{vidal2000entanglement}%
  \BibitemOpen
  \bibfield  {author} {\bibinfo {author} {\bibfnamefont {G.}~\bibnamefont {Vidal}},\ }\bibfield  {title} {\bibinfo {title} {Entanglement monotones},\ }\href {https://doi.org/10.1080/09500340008244048} {\bibfield  {journal} {\bibinfo  {journal} {Journal of Modern Optics}\ }\textbf {\bibinfo {volume} {47}},\ \bibinfo {pages} {355} (\bibinfo {year} {2000})}\BibitemShut {NoStop}%
\bibitem [{\citenamefont {Bennett}\ \emph {et~al.}(1996)\citenamefont {Bennett}, \citenamefont {DiVincenzo}, \citenamefont {Smolin},\ and\ \citenamefont {Wootters}}]{PhysRevA.54.3824}%
  \BibitemOpen
  \bibfield  {author} {\bibinfo {author} {\bibfnamefont {C.~H.}\ \bibnamefont {Bennett}}, \bibinfo {author} {\bibfnamefont {D.~P.}\ \bibnamefont {DiVincenzo}}, \bibinfo {author} {\bibfnamefont {J.~A.}\ \bibnamefont {Smolin}},\ and\ \bibinfo {author} {\bibfnamefont {W.~K.}\ \bibnamefont {Wootters}},\ }\bibfield  {title} {\bibinfo {title} {Mixed-state entanglement and quantum error correction},\ }\href {https://doi.org/10.1103/PhysRevA.54.3824} {\bibfield  {journal} {\bibinfo  {journal} {Phys. Rev. A}\ }\textbf {\bibinfo {volume} {54}},\ \bibinfo {pages} {3824} (\bibinfo {year} {1996})}\BibitemShut {NoStop}%
\bibitem [{\citenamefont {Uhlmann}(1998)}]{uhlmann1998entropy}%
  \BibitemOpen
  \bibfield  {author} {\bibinfo {author} {\bibfnamefont {A.}~\bibnamefont {Uhlmann}},\ }\bibfield  {title} {\bibinfo {title} {Entropy and optimal decompositions of states relative to a maximal commutative subalgebra},\ }\href@noop {} {\bibfield  {journal} {\bibinfo  {journal} {Open Systems \& Information Dynamics}\ }\textbf {\bibinfo {volume} {5}},\ \bibinfo {pages} {209} (\bibinfo {year} {1998})}\BibitemShut {NoStop}%
\bibitem [{\citenamefont {Wei}\ and\ \citenamefont {Goldbart}(2003)}]{wei2003geometric}%
  \BibitemOpen
  \bibfield  {author} {\bibinfo {author} {\bibfnamefont {T.-C.}\ \bibnamefont {Wei}}\ and\ \bibinfo {author} {\bibfnamefont {P.~M.}\ \bibnamefont {Goldbart}},\ }\bibfield  {title} {\bibinfo {title} {Geometric measure of entanglement and applications to bipartite and multipartite quantum states},\ }\href {https://doi.org/10.1103/PhysRevA.68.042307} {\bibfield  {journal} {\bibinfo  {journal} {Phys. Rev. A}\ }\textbf {\bibinfo {volume} {68}},\ \bibinfo {pages} {042307} (\bibinfo {year} {2003})}\BibitemShut {NoStop}%
\bibitem [{\citenamefont {Uhlmann}(2010)}]{uhlmann2010roofs}%
  \BibitemOpen
  \bibfield  {author} {\bibinfo {author} {\bibfnamefont {A.}~\bibnamefont {Uhlmann}},\ }\bibfield  {title} {\bibinfo {title} {Roofs and convexity},\ }\href {https://doi.org/10.3390/e12071799} {\bibfield  {journal} {\bibinfo  {journal} {Entropy}\ }\textbf {\bibinfo {volume} {12}},\ \bibinfo {pages} {1799} (\bibinfo {year} {2010})}\BibitemShut {NoStop}%
\bibitem [{\citenamefont {Huang}(2014)}]{huang2014computing}%
  \BibitemOpen
  \bibfield  {author} {\bibinfo {author} {\bibfnamefont {Y.}~\bibnamefont {Huang}},\ }\bibfield  {title} {\bibinfo {title} {Computing quantum discord is np-complete},\ }\href {https://doi.org/10.1088/1367-2630/16/3/033027} {\bibfield  {journal} {\bibinfo  {journal} {New Journal of Physics}\ }\textbf {\bibinfo {volume} {16}},\ \bibinfo {pages} {033027} (\bibinfo {year} {2014})}\BibitemShut {NoStop}%
\bibitem [{\citenamefont {Horodecki}(2001)}]{horodecki2001entanglement}%
  \BibitemOpen
  \bibfield  {author} {\bibinfo {author} {\bibfnamefont {M.}~\bibnamefont {Horodecki}},\ }\bibfield  {title} {\bibinfo {title} {Entanglement measures},\ }\href {https://dl.acm.org/doi/abs/10.5555/2011326.2011328} {\bibfield  {journal} {\bibinfo  {journal} {Quantum Info. Comput.}\ }\textbf {\bibinfo {volume} {1}},\ \bibinfo {pages} {3} (\bibinfo {year} {2001})}\BibitemShut {NoStop}%
\bibitem [{\citenamefont {T\'oth}\ \emph {et~al.}(2015)\citenamefont {T\'oth}, \citenamefont {Moroder},\ and\ \citenamefont {G\"uhne}}]{toth2015evaluating}%
  \BibitemOpen
  \bibfield  {author} {\bibinfo {author} {\bibfnamefont {G.}~\bibnamefont {T\'oth}}, \bibinfo {author} {\bibfnamefont {T.}~\bibnamefont {Moroder}},\ and\ \bibinfo {author} {\bibfnamefont {O.}~\bibnamefont {G\"uhne}},\ }\bibfield  {title} {\bibinfo {title} {Evaluating convex roof entanglement measures},\ }\href {https://doi.org/10.1103/PhysRevLett.114.160501} {\bibfield  {journal} {\bibinfo  {journal} {Phys. Rev. Lett.}\ }\textbf {\bibinfo {volume} {114}},\ \bibinfo {pages} {160501} (\bibinfo {year} {2015})}\BibitemShut {NoStop}%
\bibitem [{\citenamefont {Zhang}\ \emph {et~al.}(2020)\citenamefont {Zhang}, \citenamefont {Dai}, \citenamefont {Dong},\ and\ \citenamefont {Zhang}}]{zhang2020numerical}%
  \BibitemOpen
  \bibfield  {author} {\bibinfo {author} {\bibfnamefont {Z.}~\bibnamefont {Zhang}}, \bibinfo {author} {\bibfnamefont {Y.}~\bibnamefont {Dai}}, \bibinfo {author} {\bibfnamefont {Y.-L.}\ \bibnamefont {Dong}},\ and\ \bibinfo {author} {\bibfnamefont {C.}~\bibnamefont {Zhang}},\ }\bibfield  {title} {\bibinfo {title} {Numerical and analytical results for geometric measure of coherence and geometric measure of entanglement},\ }\href {http://dx.doi.org/10.1038/s41598-020-68979-z} {\bibfield  {journal} {\bibinfo  {journal} {Scientific Reports}\ }\textbf {\bibinfo {volume} {10}} (\bibinfo {year} {2020})}\BibitemShut {NoStop}%
\bibitem [{\citenamefont {Vandenberghe}\ and\ \citenamefont {Boyd}(1996)}]{vandenberghe1996semidefinite}%
  \BibitemOpen
  \bibfield  {author} {\bibinfo {author} {\bibfnamefont {L.}~\bibnamefont {Vandenberghe}}\ and\ \bibinfo {author} {\bibfnamefont {S.}~\bibnamefont {Boyd}},\ }\bibfield  {title} {\bibinfo {title} {Semidefinite programming},\ }\href {https://doi.org/10.1137/1038003} {\bibfield  {journal} {\bibinfo  {journal} {SIAM Review}\ }\textbf {\bibinfo {volume} {38}},\ \bibinfo {pages} {49} (\bibinfo {year} {1996})}\BibitemShut {NoStop}%
\bibitem [{\citenamefont {Horodecki}\ \emph {et~al.}(1998)\citenamefont {Horodecki}, \citenamefont {Horodecki},\ and\ \citenamefont {Horodecki}}]{horodecki1998mixed}%
  \BibitemOpen
  \bibfield  {author} {\bibinfo {author} {\bibfnamefont {M.}~\bibnamefont {Horodecki}}, \bibinfo {author} {\bibfnamefont {P.}~\bibnamefont {Horodecki}},\ and\ \bibinfo {author} {\bibfnamefont {R.}~\bibnamefont {Horodecki}},\ }\bibfield  {title} {\bibinfo {title} {Mixed-state entanglement and distillation: Is there a ``bound'' entanglement in nature?},\ }\href {https://doi.org/10.1103/PhysRevLett.80.5239} {\bibfield  {journal} {\bibinfo  {journal} {Phys. Rev. Lett.}\ }\textbf {\bibinfo {volume} {80}},\ \bibinfo {pages} {5239} (\bibinfo {year} {1998})}\BibitemShut {NoStop}%
\bibitem [{\citenamefont {Audenaert}\ \emph {et~al.}(2001)\citenamefont {Audenaert}, \citenamefont {Verstraete},\ and\ \citenamefont {De~Moor}}]{audenaert2001variational}%
  \BibitemOpen
  \bibfield  {author} {\bibinfo {author} {\bibfnamefont {K.}~\bibnamefont {Audenaert}}, \bibinfo {author} {\bibfnamefont {F.}~\bibnamefont {Verstraete}},\ and\ \bibinfo {author} {\bibfnamefont {B.}~\bibnamefont {De~Moor}},\ }\bibfield  {title} {\bibinfo {title} {Variational characterizations of separability and entanglement of formation},\ }\href {https://doi.org/10.1103/PhysRevA.64.052304} {\bibfield  {journal} {\bibinfo  {journal} {Phys. Rev. A}\ }\textbf {\bibinfo {volume} {64}},\ \bibinfo {pages} {052304} (\bibinfo {year} {2001})}\BibitemShut {NoStop}%
\bibitem [{\citenamefont {R\"othlisberger}\ \emph {et~al.}(2009)\citenamefont {R\"othlisberger}, \citenamefont {Lehmann},\ and\ \citenamefont {Loss}}]{rothlisberger2009numerical}%
  \BibitemOpen
  \bibfield  {author} {\bibinfo {author} {\bibfnamefont {B.}~\bibnamefont {R\"othlisberger}}, \bibinfo {author} {\bibfnamefont {J.}~\bibnamefont {Lehmann}},\ and\ \bibinfo {author} {\bibfnamefont {D.}~\bibnamefont {Loss}},\ }\bibfield  {title} {\bibinfo {title} {Numerical evaluation of convex-roof entanglement measures with applications to spin rings},\ }\href {https://doi.org/10.1103/PhysRevA.80.042301} {\bibfield  {journal} {\bibinfo  {journal} {Phys. Rev. A}\ }\textbf {\bibinfo {volume} {80}},\ \bibinfo {pages} {042301} (\bibinfo {year} {2009})}\BibitemShut {NoStop}%
\bibitem [{\citenamefont {Streltsov}\ \emph {et~al.}(2011)\citenamefont {Streltsov}, \citenamefont {Kampermann},\ and\ \citenamefont {Bru\ss{}}}]{streltsov2011simple}%
  \BibitemOpen
  \bibfield  {author} {\bibinfo {author} {\bibfnamefont {A.}~\bibnamefont {Streltsov}}, \bibinfo {author} {\bibfnamefont {H.}~\bibnamefont {Kampermann}},\ and\ \bibinfo {author} {\bibfnamefont {D.}~\bibnamefont {Bru\ss{}}},\ }\bibfield  {title} {\bibinfo {title} {Simple algorithm for computing the geometric measure of entanglement},\ }\href {https://doi.org/10.1103/PhysRevA.84.022323} {\bibfield  {journal} {\bibinfo  {journal} {Phys. Rev. A}\ }\textbf {\bibinfo {volume} {84}},\ \bibinfo {pages} {022323} (\bibinfo {year} {2011})}\BibitemShut {NoStop}%
\bibitem [{\citenamefont {Lezcano~Casado}(2019)}]{lezcano2019trivializations}%
  \BibitemOpen
  \bibfield  {author} {\bibinfo {author} {\bibfnamefont {M.}~\bibnamefont {Lezcano~Casado}},\ }\bibfield  {title} {\bibinfo {title} {Trivializations for gradient-based optimization on manifolds},\ }in\ \href {https://proceedings.neurips.cc/paper_files/paper/2019/file/1b33d16fc562464579b7199ca3114982-Paper.pdf} {\emph {\bibinfo {booktitle} {Advances in Neural Information Processing Systems}}},\ Vol.~\bibinfo {volume} {32}\ (\bibinfo  {publisher} {Curran Associates, Inc.},\ \bibinfo {year} {2019})\BibitemShut {NoStop}%
\bibitem [{\citenamefont {Manton}(2002)}]{manton2002optimization}%
  \BibitemOpen
  \bibfield  {author} {\bibinfo {author} {\bibfnamefont {J.}~\bibnamefont {Manton}},\ }\bibfield  {title} {\bibinfo {title} {Optimization algorithms exploiting unitary constraints},\ }\href {https://doi.org/10.1109/78.984753} {\bibfield  {journal} {\bibinfo  {journal} {IEEE Transactions on Signal Processing}\ }\textbf {\bibinfo {volume} {50}},\ \bibinfo {pages} {635} (\bibinfo {year} {2002})}\BibitemShut {NoStop}%
\bibitem [{\citenamefont {Paszke}\ \emph {et~al.}(2019)\citenamefont {Paszke}, \citenamefont {Gross}, \citenamefont {Massa}, \citenamefont {Lerer}, \citenamefont {Bradbury}, \citenamefont {Chanan}, \citenamefont {Killeen}, \citenamefont {Lin}, \citenamefont {Gimelshein}, \citenamefont {Antiga}, \citenamefont {Desmaison}, \citenamefont {Kopf}, \citenamefont {Yang}, \citenamefont {DeVito}, \citenamefont {Raison}, \citenamefont {Tejani}, \citenamefont {Chilamkurthy}, \citenamefont {Steiner}, \citenamefont {Fang}, \citenamefont {Bai},\ and\ \citenamefont {Chintala}}]{pytorch}%
  \BibitemOpen
  \bibfield  {author} {\bibinfo {author} {\bibfnamefont {A.}~\bibnamefont {Paszke}}, \bibinfo {author} {\bibfnamefont {S.}~\bibnamefont {Gross}}, \bibinfo {author} {\bibfnamefont {F.}~\bibnamefont {Massa}}, \bibinfo {author} {\bibfnamefont {A.}~\bibnamefont {Lerer}}, \bibinfo {author} {\bibfnamefont {J.}~\bibnamefont {Bradbury}}, \bibinfo {author} {\bibfnamefont {G.}~\bibnamefont {Chanan}}, \bibinfo {author} {\bibfnamefont {T.}~\bibnamefont {Killeen}}, \bibinfo {author} {\bibfnamefont {Z.}~\bibnamefont {Lin}}, \bibinfo {author} {\bibfnamefont {N.}~\bibnamefont {Gimelshein}}, \bibinfo {author} {\bibfnamefont {L.}~\bibnamefont {Antiga}}, \bibinfo {author} {\bibfnamefont {A.}~\bibnamefont {Desmaison}}, \bibinfo {author} {\bibfnamefont {A.}~\bibnamefont {Kopf}}, \bibinfo {author} {\bibfnamefont {E.}~\bibnamefont {Yang}}, \bibinfo {author} {\bibfnamefont {Z.}~\bibnamefont {DeVito}}, \bibinfo {author} {\bibfnamefont {M.}~\bibnamefont {Raison}}, \bibinfo {author} {\bibfnamefont {A.}~\bibnamefont {Tejani}}, \bibinfo
  {author} {\bibfnamefont {S.}~\bibnamefont {Chilamkurthy}}, \bibinfo {author} {\bibfnamefont {B.}~\bibnamefont {Steiner}}, \bibinfo {author} {\bibfnamefont {L.}~\bibnamefont {Fang}}, \bibinfo {author} {\bibfnamefont {J.}~\bibnamefont {Bai}},\ and\ \bibinfo {author} {\bibfnamefont {S.}~\bibnamefont {Chintala}},\ }\bibfield  {title} {\bibinfo {title} {Pytorch: An imperative style, high-performance deep learning library},\ }in\ \href {http://papers.neurips.cc/paper/9015-pytorch-an-imperative-style-high-performance-deep-learning-library.pdf} {\emph {\bibinfo {booktitle} {Advances in Neural Information Processing Systems 32}}},\ \bibinfo {editor} {edited by\ \bibinfo {editor} {\bibfnamefont {H.}~\bibnamefont {Wallach}}, \bibinfo {editor} {\bibfnamefont {H.}~\bibnamefont {Larochelle}}, \bibinfo {editor} {\bibfnamefont {A.}~\bibnamefont {Beygelzimer}}, \bibinfo {editor} {\bibfnamefont {F.}~\bibnamefont {d\textquotesingle Alch\'{e}-Buc}}, \bibinfo {editor} {\bibfnamefont {E.}~\bibnamefont {Fox}},\ and\ \bibinfo
  {editor} {\bibfnamefont {R.}~\bibnamefont {Garnett}}}\ (\bibinfo  {publisher} {Curran Associates, Inc.},\ \bibinfo {year} {2019})\ pp.\ \bibinfo {pages} {8024--8035}\BibitemShut {NoStop}%
\bibitem [{\citenamefont {Wootters}(2001)}]{wootters2001entanglement}%
  \BibitemOpen
  \bibfield  {author} {\bibinfo {author} {\bibfnamefont {W.~K.}\ \bibnamefont {Wootters}},\ }\bibfield  {title} {\bibinfo {title} {Entanglement of formation and concurrence},\ }\href {https://dl.acm.org/doi/10.5555/2011326.2011329} {\bibfield  {journal} {\bibinfo  {journal} {Quantum Info. Comput.}\ }\textbf {\bibinfo {volume} {1}},\ \bibinfo {pages} {27–44} (\bibinfo {year} {2001})}\BibitemShut {NoStop}%
\bibitem [{\citenamefont {Buscemi}\ \emph {et~al.}(2007)\citenamefont {Buscemi}, \citenamefont {Bordone},\ and\ \citenamefont {Bertoni}}]{buscemi2007linear}%
  \BibitemOpen
  \bibfield  {author} {\bibinfo {author} {\bibfnamefont {F.}~\bibnamefont {Buscemi}}, \bibinfo {author} {\bibfnamefont {P.}~\bibnamefont {Bordone}},\ and\ \bibinfo {author} {\bibfnamefont {A.}~\bibnamefont {Bertoni}},\ }\bibfield  {title} {\bibinfo {title} {Linear entropy as an entanglement measure in two-fermion systems},\ }\href {https://doi.org/10.1103/PhysRevA.75.032301} {\bibfield  {journal} {\bibinfo  {journal} {Phys. Rev. A}\ }\textbf {\bibinfo {volume} {75}},\ \bibinfo {pages} {032301} (\bibinfo {year} {2007})}\BibitemShut {NoStop}%
\bibitem [{\citenamefont {Streltsov}\ \emph {et~al.}(2015)\citenamefont {Streltsov}, \citenamefont {Singh}, \citenamefont {Dhar}, \citenamefont {Bera},\ and\ \citenamefont {Adesso}}]{PhysRevLett.115.020403}%
  \BibitemOpen
  \bibfield  {author} {\bibinfo {author} {\bibfnamefont {A.}~\bibnamefont {Streltsov}}, \bibinfo {author} {\bibfnamefont {U.}~\bibnamefont {Singh}}, \bibinfo {author} {\bibfnamefont {H.~S.}\ \bibnamefont {Dhar}}, \bibinfo {author} {\bibfnamefont {M.~N.}\ \bibnamefont {Bera}},\ and\ \bibinfo {author} {\bibfnamefont {G.}~\bibnamefont {Adesso}},\ }\bibfield  {title} {\bibinfo {title} {Measuring quantum coherence with entanglement},\ }\href {https://doi.org/10.1103/PhysRevLett.115.020403} {\bibfield  {journal} {\bibinfo  {journal} {Phys. Rev. Lett.}\ }\textbf {\bibinfo {volume} {115}},\ \bibinfo {pages} {020403} (\bibinfo {year} {2015})}\BibitemShut {NoStop}%
\bibitem [{\citenamefont {Leone}\ \emph {et~al.}(2022)\citenamefont {Leone}, \citenamefont {Oliviero},\ and\ \citenamefont {Hamma}}]{leone2022stabilizer}%
  \BibitemOpen
  \bibfield  {author} {\bibinfo {author} {\bibfnamefont {L.}~\bibnamefont {Leone}}, \bibinfo {author} {\bibfnamefont {S.~F.~E.}\ \bibnamefont {Oliviero}},\ and\ \bibinfo {author} {\bibfnamefont {A.}~\bibnamefont {Hamma}},\ }\bibfield  {title} {\bibinfo {title} {Stabilizer r\'enyi entropy},\ }\href {https://doi.org/10.1103/PhysRevLett.128.050402} {\bibfield  {journal} {\bibinfo  {journal} {Phys. Rev. Lett.}\ }\textbf {\bibinfo {volume} {128}},\ \bibinfo {pages} {050402} (\bibinfo {year} {2022})}\BibitemShut {NoStop}%
\bibitem [{\citenamefont {Haug}\ and\ \citenamefont {Piroli}(2023{\natexlab{a}})}]{haug2023stabilizer}%
  \BibitemOpen
  \bibfield  {author} {\bibinfo {author} {\bibfnamefont {T.}~\bibnamefont {Haug}}\ and\ \bibinfo {author} {\bibfnamefont {L.}~\bibnamefont {Piroli}},\ }\bibfield  {title} {\bibinfo {title} {Stabilizer entropies and nonstabilizerness monotones},\ }\href {https://doi.org/10.22331/q-2023-08-28-1092} {\bibfield  {journal} {\bibinfo  {journal} {{Quantum}}\ }\textbf {\bibinfo {volume} {7}},\ \bibinfo {pages} {1092} (\bibinfo {year} {2023}{\natexlab{a}})}\BibitemShut {NoStop}%
\bibitem [{\citenamefont {Leone}\ and\ \citenamefont {Bittel}(2024)}]{leone2024stabilizer}%
  \BibitemOpen
  \bibfield  {author} {\bibinfo {author} {\bibfnamefont {L.}~\bibnamefont {Leone}}\ and\ \bibinfo {author} {\bibfnamefont {L.}~\bibnamefont {Bittel}},\ }\bibfield  {title} {\bibinfo {title} {Stabilizer entropies are monotones for magic-state resource theory},\ }\href@noop {} {\bibfield  {journal} {\bibinfo  {journal} {arXiv preprint arXiv:2404.11652}\ } (\bibinfo {year} {2024})}\BibitemShut {NoStop}%
\bibitem [{\citenamefont {Shor}(2003)}]{shor2003capacities}%
  \BibitemOpen
  \bibfield  {author} {\bibinfo {author} {\bibfnamefont {P.~W.}\ \bibnamefont {Shor}},\ }\bibfield  {title} {\bibinfo {title} {Capacities of quantum channels and how to find them},\ }\href {https://doi.org/10.1007/s10107-003-0446-y} {\bibfield  {journal} {\bibinfo  {journal} {Mathematical Programming}\ }\textbf {\bibinfo {volume} {97}},\ \bibinfo {pages} {311–335} (\bibinfo {year} {2003})}\BibitemShut {NoStop}%
\bibitem [{\citenamefont {Shor}(2004)}]{shor2004equivalence}%
  \BibitemOpen
  \bibfield  {author} {\bibinfo {author} {\bibfnamefont {P.~W.}\ \bibnamefont {Shor}},\ }\bibfield  {title} {\bibinfo {title} {Equivalence of additivity questions in quantum information theory},\ }\href {https://doi.org/10.1007/s00220-003-0981-7} {\bibfield  {journal} {\bibinfo  {journal} {Communications in Mathematical Physics}\ }\textbf {\bibinfo {volume} {246}},\ \bibinfo {pages} {453–472} (\bibinfo {year} {2004})}\BibitemShut {NoStop}%
\bibitem [{\citenamefont {Holevo}(2011)}]{holevo2011probabilistic}%
  \BibitemOpen
  \bibfield  {author} {\bibinfo {author} {\bibfnamefont {A.}~\bibnamefont {Holevo}},\ }\href {https://doi.org/10.1007/978-88-7642-378-9} {\emph {\bibinfo {title} {Probabilistic and Statistical Aspects of Quantum Theory}}}\ (\bibinfo  {publisher} {Edizioni della Normale},\ \bibinfo {year} {2011})\BibitemShut {NoStop}%
\bibitem [{\citenamefont {Streltsov}\ \emph {et~al.}(2010)\citenamefont {Streltsov}, \citenamefont {Kampermann},\ and\ \citenamefont {Bruß}}]{streltsov2010linking}%
  \BibitemOpen
  \bibfield  {author} {\bibinfo {author} {\bibfnamefont {A.}~\bibnamefont {Streltsov}}, \bibinfo {author} {\bibfnamefont {H.}~\bibnamefont {Kampermann}},\ and\ \bibinfo {author} {\bibfnamefont {D.}~\bibnamefont {Bruß}},\ }\bibfield  {title} {\bibinfo {title} {Linking a distance measure of entanglement to its convex roof},\ }\href {https://doi.org/10.1088/1367-2630/12/12/123004} {\bibfield  {journal} {\bibinfo  {journal} {New Journal of Physics}\ }\textbf {\bibinfo {volume} {12}},\ \bibinfo {pages} {123004} (\bibinfo {year} {2010})}\BibitemShut {NoStop}%
\bibitem [{\citenamefont {Lockhart}(2000)}]{lockhart2000optimal}%
  \BibitemOpen
  \bibfield  {author} {\bibinfo {author} {\bibfnamefont {R.}~\bibnamefont {Lockhart}},\ }\bibfield  {title} {\bibinfo {title} {Optimal ensemble length of mixed separable states},\ }\href {https://doi.org/10.1063/1.1290055} {\bibfield  {journal} {\bibinfo  {journal} {Journal of Mathematical Physics}\ }\textbf {\bibinfo {volume} {41}},\ \bibinfo {pages} {6766} (\bibinfo {year} {2000})}\BibitemShut {NoStop}%
\bibitem [{\citenamefont {Wootters}(1998)}]{wootters1998entanglement}%
  \BibitemOpen
  \bibfield  {author} {\bibinfo {author} {\bibfnamefont {W.~K.}\ \bibnamefont {Wootters}},\ }\bibfield  {title} {\bibinfo {title} {Entanglement of formation of an arbitrary state of two qubits},\ }\href {https://doi.org/10.1103/PhysRevLett.80.2245} {\bibfield  {journal} {\bibinfo  {journal} {Phys. Rev. Lett.}\ }\textbf {\bibinfo {volume} {80}},\ \bibinfo {pages} {2245} (\bibinfo {year} {1998})}\BibitemShut {NoStop}%
\bibitem [{\citenamefont {Chen}\ \emph {et~al.}(2022)\citenamefont {Chen}, \citenamefont {He},\ and\ \citenamefont {Zhang}}]{chen2022tight}%
  \BibitemOpen
  \bibfield  {author} {\bibinfo {author} {\bibfnamefont {X.}~\bibnamefont {Chen}}, \bibinfo {author} {\bibfnamefont {Y.}~\bibnamefont {He}},\ and\ \bibinfo {author} {\bibfnamefont {Z.}~\bibnamefont {Zhang}},\ }\bibfield  {title} {\bibinfo {title} {Tight error bounds for nonnegative orthogonality constraints and exact penalties},\ }\href@noop {} {\bibfield  {journal} {\bibinfo  {journal} {arXiv preprint arXiv:2210.05164}\ } (\bibinfo {year} {2022})}\BibitemShut {NoStop}%
\bibitem [{sup()}]{supple}%
  \BibitemOpen
  \href@noop {} {}\bibinfo {note} {See Supplementary Material for more details.}\BibitemShut {Stop}%
\bibitem [{\citenamefont {Virtanen}\ \emph {et~al.}(2020{\natexlab{a}})\citenamefont {Virtanen}, \citenamefont {Gommers}, \citenamefont {Oliphant}, \citenamefont {Haberland}, \citenamefont {Reddy}, \citenamefont {Cournapeau}, \citenamefont {Burovski}, \citenamefont {Peterson}, \citenamefont {Weckesser}, \citenamefont {Bright} \emph {et~al.}}]{virtanen2020scipy}%
  \BibitemOpen
  \bibfield  {author} {\bibinfo {author} {\bibfnamefont {P.}~\bibnamefont {Virtanen}}, \bibinfo {author} {\bibfnamefont {R.}~\bibnamefont {Gommers}}, \bibinfo {author} {\bibfnamefont {T.~E.}\ \bibnamefont {Oliphant}}, \bibinfo {author} {\bibfnamefont {M.}~\bibnamefont {Haberland}}, \bibinfo {author} {\bibfnamefont {T.}~\bibnamefont {Reddy}}, \bibinfo {author} {\bibfnamefont {D.}~\bibnamefont {Cournapeau}}, \bibinfo {author} {\bibfnamefont {E.}~\bibnamefont {Burovski}}, \bibinfo {author} {\bibfnamefont {P.}~\bibnamefont {Peterson}}, \bibinfo {author} {\bibfnamefont {W.}~\bibnamefont {Weckesser}}, \bibinfo {author} {\bibfnamefont {J.}~\bibnamefont {Bright}}, \emph {et~al.},\ }\bibfield  {title} {\bibinfo {title} {Scipy 1.0: fundamental algorithms for scientific computing in python},\ }\href@noop {} {\bibfield  {journal} {\bibinfo  {journal} {Nature methods}\ }\textbf {\bibinfo {volume} {17}},\ \bibinfo {pages} {261} (\bibinfo {year} {2020}{\natexlab{a}})}\BibitemShut {NoStop}%
\bibitem [{\citenamefont {Zhu}\ \emph {et~al.}(1997)\citenamefont {Zhu}, \citenamefont {Byrd}, \citenamefont {Lu},\ and\ \citenamefont {Nocedal}}]{10.1145/279232.279236}%
  \BibitemOpen
  \bibfield  {author} {\bibinfo {author} {\bibfnamefont {C.}~\bibnamefont {Zhu}}, \bibinfo {author} {\bibfnamefont {R.~H.}\ \bibnamefont {Byrd}}, \bibinfo {author} {\bibfnamefont {P.}~\bibnamefont {Lu}},\ and\ \bibinfo {author} {\bibfnamefont {J.}~\bibnamefont {Nocedal}},\ }\bibfield  {title} {\bibinfo {title} {Algorithm 778: L-bfgs-b: Fortran subroutines for large-scale bound-constrained optimization},\ }\href {https://doi.org/10.1145/279232.279236} {\bibfield  {journal} {\bibinfo  {journal} {ACM Trans. Math. Softw.}\ }\textbf {\bibinfo {volume} {23}},\ \bibinfo {pages} {550} (\bibinfo {year} {1997})}\BibitemShut {NoStop}%
\bibitem [{\citenamefont {Rungta}\ and\ \citenamefont {Caves}(2003)}]{PhysRevA.67.012307}%
  \BibitemOpen
  \bibfield  {author} {\bibinfo {author} {\bibfnamefont {P.}~\bibnamefont {Rungta}}\ and\ \bibinfo {author} {\bibfnamefont {C.~M.}\ \bibnamefont {Caves}},\ }\bibfield  {title} {\bibinfo {title} {Concurrence-based entanglement measures for isotropic states},\ }\href {https://doi.org/10.1103/PhysRevA.67.012307} {\bibfield  {journal} {\bibinfo  {journal} {Phys. Rev. A}\ }\textbf {\bibinfo {volume} {67}},\ \bibinfo {pages} {012307} (\bibinfo {year} {2003})}\BibitemShut {NoStop}%
\bibitem [{\citenamefont {Gao}\ \emph {et~al.}(2008)\citenamefont {Gao}, \citenamefont {Sergio}, \citenamefont {Chen}, \citenamefont {Fei},\ and\ \citenamefont {Li-Jost}}]{gao2008entanglement}%
  \BibitemOpen
  \bibfield  {author} {\bibinfo {author} {\bibfnamefont {X.}~\bibnamefont {Gao}}, \bibinfo {author} {\bibfnamefont {A.}~\bibnamefont {Sergio}}, \bibinfo {author} {\bibfnamefont {K.}~\bibnamefont {Chen}}, \bibinfo {author} {\bibfnamefont {S.}~\bibnamefont {Fei}},\ and\ \bibinfo {author} {\bibfnamefont {X.}~\bibnamefont {Li-Jost}},\ }\bibfield  {title} {\bibinfo {title} {Entanglement of formation and concurrence for mixed states},\ }\href {https://doi.org/10.1007/s11704-008-0017-8} {\bibfield  {journal} {\bibinfo  {journal} {Frontiers of Computer Science in China}\ }\textbf {\bibinfo {volume} {2}},\ \bibinfo {pages} {114–128} (\bibinfo {year} {2008})}\BibitemShut {NoStop}%
\bibitem [{\citenamefont {Horodecki}(1997)}]{horodecki1997separability}%
  \BibitemOpen
  \bibfield  {author} {\bibinfo {author} {\bibfnamefont {P.}~\bibnamefont {Horodecki}},\ }\bibfield  {title} {\bibinfo {title} {Separability criterion and inseparable mixed states with positive partial transposition},\ }\href {https://doi.org/https://doi.org/10.1016/S0375-9601(97)00416-7} {\bibfield  {journal} {\bibinfo  {journal} {Physics Letters A}\ }\textbf {\bibinfo {volume} {232}},\ \bibinfo {pages} {333} (\bibinfo {year} {1997})}\BibitemShut {NoStop}%
\bibitem [{\citenamefont {Oliviero}\ \emph {et~al.}(2022{\natexlab{a}})\citenamefont {Oliviero}, \citenamefont {Leone},\ and\ \citenamefont {Hamma}}]{PhysRevA.106.042426}%
  \BibitemOpen
  \bibfield  {author} {\bibinfo {author} {\bibfnamefont {S.~F.~E.}\ \bibnamefont {Oliviero}}, \bibinfo {author} {\bibfnamefont {L.}~\bibnamefont {Leone}},\ and\ \bibinfo {author} {\bibfnamefont {A.}~\bibnamefont {Hamma}},\ }\bibfield  {title} {\bibinfo {title} {Magic-state resource theory for the ground state of the transverse-field ising model},\ }\href {https://doi.org/10.1103/PhysRevA.106.042426} {\bibfield  {journal} {\bibinfo  {journal} {Phys. Rev. A}\ }\textbf {\bibinfo {volume} {106}},\ \bibinfo {pages} {042426} (\bibinfo {year} {2022}{\natexlab{a}})}\BibitemShut {NoStop}%
\bibitem [{\citenamefont {Haug}\ and\ \citenamefont {Piroli}(2023{\natexlab{b}})}]{PhysRevB.107.035148}%
  \BibitemOpen
  \bibfield  {author} {\bibinfo {author} {\bibfnamefont {T.}~\bibnamefont {Haug}}\ and\ \bibinfo {author} {\bibfnamefont {L.}~\bibnamefont {Piroli}},\ }\bibfield  {title} {\bibinfo {title} {Quantifying nonstabilizerness of matrix product states},\ }\href {https://doi.org/10.1103/PhysRevB.107.035148} {\bibfield  {journal} {\bibinfo  {journal} {Phys. Rev. B}\ }\textbf {\bibinfo {volume} {107}},\ \bibinfo {pages} {035148} (\bibinfo {year} {2023}{\natexlab{b}})}\BibitemShut {NoStop}%
\bibitem [{\citenamefont {Lami}\ and\ \citenamefont {Collura}(2023)}]{PhysRevLett.131.180401}%
  \BibitemOpen
  \bibfield  {author} {\bibinfo {author} {\bibfnamefont {G.}~\bibnamefont {Lami}}\ and\ \bibinfo {author} {\bibfnamefont {M.}~\bibnamefont {Collura}},\ }\bibfield  {title} {\bibinfo {title} {Nonstabilizerness via perfect pauli sampling of matrix product states},\ }\href {https://doi.org/10.1103/PhysRevLett.131.180401} {\bibfield  {journal} {\bibinfo  {journal} {Phys. Rev. Lett.}\ }\textbf {\bibinfo {volume} {131}},\ \bibinfo {pages} {180401} (\bibinfo {year} {2023})}\BibitemShut {NoStop}%
\bibitem [{\citenamefont {Oliviero}\ \emph {et~al.}(2022{\natexlab{b}})\citenamefont {Oliviero}, \citenamefont {Leone}, \citenamefont {Hamma},\ and\ \citenamefont {Lloyd}}]{oliviero2022measuring}%
  \BibitemOpen
  \bibfield  {author} {\bibinfo {author} {\bibfnamefont {S.~F.~E.}\ \bibnamefont {Oliviero}}, \bibinfo {author} {\bibfnamefont {L.}~\bibnamefont {Leone}}, \bibinfo {author} {\bibfnamefont {A.}~\bibnamefont {Hamma}},\ and\ \bibinfo {author} {\bibfnamefont {S.}~\bibnamefont {Lloyd}},\ }\bibfield  {title} {\bibinfo {title} {Measuring magic on a quantum processor},\ }\href {https://www.nature.com/articles/s41534-022-00666-5#citeas} {\bibfield  {journal} {\bibinfo  {journal} {npj Quantum Information}\ }\textbf {\bibinfo {volume} {8}} (\bibinfo {year} {2022}{\natexlab{b}})}\BibitemShut {NoStop}%
\bibitem [{\citenamefont {Holevo}(1998)}]{holevo1998capacity}%
  \BibitemOpen
  \bibfield  {author} {\bibinfo {author} {\bibfnamefont {A.}~\bibnamefont {Holevo}},\ }\bibfield  {title} {\bibinfo {title} {The capacity of the quantum channel with general signal states},\ }\href {https://doi.org/10.1109/18.651037} {\bibfield  {journal} {\bibinfo  {journal} {IEEE Transactions on Information Theory}\ }\textbf {\bibinfo {volume} {44}},\ \bibinfo {pages} {269} (\bibinfo {year} {1998})}\BibitemShut {NoStop}%
\bibitem [{\citenamefont {Schumacher}\ and\ \citenamefont {Westmoreland}(1997)}]{schumacher1997sending}%
  \BibitemOpen
  \bibfield  {author} {\bibinfo {author} {\bibfnamefont {B.}~\bibnamefont {Schumacher}}\ and\ \bibinfo {author} {\bibfnamefont {M.~D.}\ \bibnamefont {Westmoreland}},\ }\bibfield  {title} {\bibinfo {title} {Sending classical information via noisy quantum channels},\ }\href {https://doi.org/10.1103/PhysRevA.56.131} {\bibfield  {journal} {\bibinfo  {journal} {Phys. Rev. A}\ }\textbf {\bibinfo {volume} {56}},\ \bibinfo {pages} {131} (\bibinfo {year} {1997})}\BibitemShut {NoStop}%
\bibitem [{\citenamefont {Giovannetti}\ \emph {et~al.}(2004)\citenamefont {Giovannetti}, \citenamefont {Lloyd},\ and\ \citenamefont {Maccone}}]{giovannetti2004quantum}%
  \BibitemOpen
  \bibfield  {author} {\bibinfo {author} {\bibfnamefont {V.}~\bibnamefont {Giovannetti}}, \bibinfo {author} {\bibfnamefont {S.}~\bibnamefont {Lloyd}},\ and\ \bibinfo {author} {\bibfnamefont {L.}~\bibnamefont {Maccone}},\ }\bibfield  {title} {\bibinfo {title} {Quantum-enhanced measurements: Beating the standard quantum limit},\ }\href {https://doi.org/10.1126/science.1104149} {\bibfield  {journal} {\bibinfo  {journal} {Science}\ }\textbf {\bibinfo {volume} {306}},\ \bibinfo {pages} {1330} (\bibinfo {year} {2004})}\BibitemShut {NoStop}%
\bibitem [{\citenamefont {Giovannetti}\ \emph {et~al.}(2006)\citenamefont {Giovannetti}, \citenamefont {Lloyd},\ and\ \citenamefont {Maccone}}]{giovannetti2006quantum}%
  \BibitemOpen
  \bibfield  {author} {\bibinfo {author} {\bibfnamefont {V.}~\bibnamefont {Giovannetti}}, \bibinfo {author} {\bibfnamefont {S.}~\bibnamefont {Lloyd}},\ and\ \bibinfo {author} {\bibfnamefont {L.}~\bibnamefont {Maccone}},\ }\bibfield  {title} {\bibinfo {title} {Quantum metrology},\ }\href {https://doi.org/10.1103/PhysRevLett.96.010401} {\bibfield  {journal} {\bibinfo  {journal} {Phys. Rev. Lett.}\ }\textbf {\bibinfo {volume} {96}},\ \bibinfo {pages} {010401} (\bibinfo {year} {2006})}\BibitemShut {NoStop}%
\bibitem [{\citenamefont {T\'oth}\ and\ \citenamefont {Petz}(2013)}]{toth2013extremal}%
  \BibitemOpen
  \bibfield  {author} {\bibinfo {author} {\bibfnamefont {G.}~\bibnamefont {T\'oth}}\ and\ \bibinfo {author} {\bibfnamefont {D.}~\bibnamefont {Petz}},\ }\bibfield  {title} {\bibinfo {title} {Extremal properties of the variance and the quantum fisher information},\ }\href {https://doi.org/10.1103/PhysRevA.87.032324} {\bibfield  {journal} {\bibinfo  {journal} {Phys. Rev. A}\ }\textbf {\bibinfo {volume} {87}},\ \bibinfo {pages} {032324} (\bibinfo {year} {2013})}\BibitemShut {NoStop}%
\bibitem [{\citenamefont {Yu}(2013)}]{yu2013quantum}%
  \BibitemOpen
  \bibfield  {author} {\bibinfo {author} {\bibfnamefont {S.}~\bibnamefont {Yu}},\ }\bibfield  {title} {\bibinfo {title} {Quantum fisher information as the convex roof of variance},\ }\href@noop {} {\bibfield  {journal} {\bibinfo  {journal} {arXiv preprint arXiv:1302.5311}\ } (\bibinfo {year} {2013})}\BibitemShut {NoStop}%
\bibitem [{\citenamefont {Zhao}\ \emph {et~al.}(2022)\citenamefont {Zhao}, \citenamefont {Zhang},\ and\ \citenamefont {Fei}}]{zhao2022standard}%
  \BibitemOpen
  \bibfield  {author} {\bibinfo {author} {\bibfnamefont {M.-J.}\ \bibnamefont {Zhao}}, \bibinfo {author} {\bibfnamefont {L.}~\bibnamefont {Zhang}},\ and\ \bibinfo {author} {\bibfnamefont {S.-M.}\ \bibnamefont {Fei}},\ }\bibfield  {title} {\bibinfo {title} {Standard symmetrized variance with applications to coherence, uncertainty, and entanglement},\ }\href {https://doi.org/10.1103/PhysRevA.106.012417} {\bibfield  {journal} {\bibinfo  {journal} {Phys. Rev. A}\ }\textbf {\bibinfo {volume} {106}},\ \bibinfo {pages} {012417} (\bibinfo {year} {2022})}\BibitemShut {NoStop}%
\bibitem [{\citenamefont {Zhu}\ and\ \citenamefont {Zhang}(2024)}]{zhu_2024_12540301}%
  \BibitemOpen
  \bibfield  {author} {\bibinfo {author} {\bibfnamefont {X.}~\bibnamefont {Zhu}}\ and\ \bibinfo {author} {\bibfnamefont {C.}~\bibnamefont {Zhang}},\ }\href {https://doi.org/10.5281/zenodo.12540301} {\bibinfo {title} {{numqi/dm-stiefel: convex roof extension for various quantum resources}}} (\bibinfo {year} {2024})\BibitemShut {NoStop}%
\bibitem [{\citenamefont {Bertlmann}\ and\ \citenamefont {Krammer}(2008)}]{bertlmann2008bloch}%
  \BibitemOpen
  \bibfield  {author} {\bibinfo {author} {\bibfnamefont {R.~A.}\ \bibnamefont {Bertlmann}}\ and\ \bibinfo {author} {\bibfnamefont {P.}~\bibnamefont {Krammer}},\ }\bibfield  {title} {\bibinfo {title} {Bloch vectors for qudits},\ }\href@noop {} {\bibfield  {journal} {\bibinfo  {journal} {Journal of Physics A: Mathematical and Theoretical}\ }\textbf {\bibinfo {volume} {41}},\ \bibinfo {pages} {235303} (\bibinfo {year} {2008})}\BibitemShut {NoStop}%
\bibitem [{\citenamefont {Lezcano-Casado}\ and\ \citenamefont {Mart\'{\i}nez-Rubio}(2019)}]{pmlr-v97-lezcano-casado19a}%
  \BibitemOpen
  \bibfield  {author} {\bibinfo {author} {\bibfnamefont {M.}~\bibnamefont {Lezcano-Casado}}\ and\ \bibinfo {author} {\bibfnamefont {D.}~\bibnamefont {Mart\'{\i}nez-Rubio}},\ }\bibfield  {title} {\bibinfo {title} {Cheap orthogonal constraints in neural networks: A simple parametrization of the orthogonal and unitary group},\ }in\ \href {https://proceedings.mlr.press/v97/lezcano-casado19a.html} {\emph {\bibinfo {booktitle} {Proceedings of the 36th International Conference on Machine Learning}}},\ \bibinfo {series} {Proceedings of Machine Learning Research}, Vol.~\bibinfo {volume} {97},\ \bibinfo {editor} {edited by\ \bibinfo {editor} {\bibfnamefont {K.}~\bibnamefont {Chaudhuri}}\ and\ \bibinfo {editor} {\bibfnamefont {R.}~\bibnamefont {Salakhutdinov}}}\ (\bibinfo  {publisher} {PMLR},\ \bibinfo {year} {2019})\ pp.\ \bibinfo {pages} {3794--3803}\BibitemShut {NoStop}%
\bibitem [{\citenamefont {Cardoso}(2012)}]{cardoso2012computation}%
  \BibitemOpen
  \bibfield  {author} {\bibinfo {author} {\bibfnamefont {J.~R.}\ \bibnamefont {Cardoso}},\ }\bibfield  {title} {\bibinfo {title} {Computation of the matrix pth root and its fr{\'e}chet derivative by integrals},\ }\href@noop {} {\bibfield  {journal} {\bibinfo  {journal} {Electron. Trans. Numer. Anal}\ }\textbf {\bibinfo {volume} {39}},\ \bibinfo {pages} {414} (\bibinfo {year} {2012})}\BibitemShut {NoStop}%
\bibitem [{\citenamefont {Virtanen}\ \emph {et~al.}(2020{\natexlab{b}})\citenamefont {Virtanen}, \citenamefont {Gommers}, \citenamefont {Oliphant}, \citenamefont {Haberland}, \citenamefont {Reddy}, \citenamefont {Cournapeau}, \citenamefont {Burovski}, \citenamefont {Peterson}, \citenamefont {Weckesser}, \citenamefont {Bright}, \citenamefont {{van der Walt}}, \citenamefont {Brett}, \citenamefont {Wilson}, \citenamefont {Millman}, \citenamefont {Mayorov}, \citenamefont {Nelson}, \citenamefont {Jones}, \citenamefont {Kern}, \citenamefont {Larson}, \citenamefont {Carey}, \citenamefont {Polat}, \citenamefont {Feng}, \citenamefont {Moore}, \citenamefont {{VanderPlas}}, \citenamefont {Laxalde}, \citenamefont {Perktold}, \citenamefont {Cimrman}, \citenamefont {Henriksen}, \citenamefont {Quintero}, \citenamefont {Harris}, \citenamefont {Archibald}, \citenamefont {Ribeiro}, \citenamefont {Pedregosa}, \citenamefont {{van Mulbregt}},\ and\ \citenamefont {{SciPy 1.0 Contributors}}}]{2020SciPy-NMeth}%
  \BibitemOpen
  \bibfield  {author} {\bibinfo {author} {\bibfnamefont {P.}~\bibnamefont {Virtanen}}, \bibinfo {author} {\bibfnamefont {R.}~\bibnamefont {Gommers}}, \bibinfo {author} {\bibfnamefont {T.~E.}\ \bibnamefont {Oliphant}}, \bibinfo {author} {\bibfnamefont {M.}~\bibnamefont {Haberland}}, \bibinfo {author} {\bibfnamefont {T.}~\bibnamefont {Reddy}}, \bibinfo {author} {\bibfnamefont {D.}~\bibnamefont {Cournapeau}}, \bibinfo {author} {\bibfnamefont {E.}~\bibnamefont {Burovski}}, \bibinfo {author} {\bibfnamefont {P.}~\bibnamefont {Peterson}}, \bibinfo {author} {\bibfnamefont {W.}~\bibnamefont {Weckesser}}, \bibinfo {author} {\bibfnamefont {J.}~\bibnamefont {Bright}}, \bibinfo {author} {\bibfnamefont {S.~J.}\ \bibnamefont {{van der Walt}}}, \bibinfo {author} {\bibfnamefont {M.}~\bibnamefont {Brett}}, \bibinfo {author} {\bibfnamefont {J.}~\bibnamefont {Wilson}}, \bibinfo {author} {\bibfnamefont {K.~J.}\ \bibnamefont {Millman}}, \bibinfo {author} {\bibfnamefont {N.}~\bibnamefont {Mayorov}}, \bibinfo {author} {\bibfnamefont
  {A.~R.~J.}\ \bibnamefont {Nelson}}, \bibinfo {author} {\bibfnamefont {E.}~\bibnamefont {Jones}}, \bibinfo {author} {\bibfnamefont {R.}~\bibnamefont {Kern}}, \bibinfo {author} {\bibfnamefont {E.}~\bibnamefont {Larson}}, \bibinfo {author} {\bibfnamefont {C.~J.}\ \bibnamefont {Carey}}, \bibinfo {author} {\bibfnamefont {{\.I}.}~\bibnamefont {Polat}}, \bibinfo {author} {\bibfnamefont {Y.}~\bibnamefont {Feng}}, \bibinfo {author} {\bibfnamefont {E.~W.}\ \bibnamefont {Moore}}, \bibinfo {author} {\bibfnamefont {J.}~\bibnamefont {{VanderPlas}}}, \bibinfo {author} {\bibfnamefont {D.}~\bibnamefont {Laxalde}}, \bibinfo {author} {\bibfnamefont {J.}~\bibnamefont {Perktold}}, \bibinfo {author} {\bibfnamefont {R.}~\bibnamefont {Cimrman}}, \bibinfo {author} {\bibfnamefont {I.}~\bibnamefont {Henriksen}}, \bibinfo {author} {\bibfnamefont {E.~A.}\ \bibnamefont {Quintero}}, \bibinfo {author} {\bibfnamefont {C.~R.}\ \bibnamefont {Harris}}, \bibinfo {author} {\bibfnamefont {A.~M.}\ \bibnamefont {Archibald}}, \bibinfo {author}
  {\bibfnamefont {A.~H.}\ \bibnamefont {Ribeiro}}, \bibinfo {author} {\bibfnamefont {F.}~\bibnamefont {Pedregosa}}, \bibinfo {author} {\bibfnamefont {P.}~\bibnamefont {{van Mulbregt}}},\ and\ \bibinfo {author} {\bibnamefont {{SciPy 1.0 Contributors}}},\ }\bibfield  {title} {\bibinfo {title} {{{SciPy} 1.0: Fundamental Algorithms for Scientific Computing in Python}},\ }\href {https://doi.org/10.1038/s41592-019-0686-2} {\bibfield  {journal} {\bibinfo  {journal} {Nature Methods}\ }\textbf {\bibinfo {volume} {17}},\ \bibinfo {pages} {261} (\bibinfo {year} {2020}{\natexlab{b}})}\BibitemShut {NoStop}%
\bibitem [{\citenamefont {a.~Smith}\ and\ \citenamefont {Gray}(2018)}]{Smith_opt_einsum}%
  \BibitemOpen
  \bibfield  {author} {\bibinfo {author} {\bibfnamefont {D.~G.}\ \bibnamefont {a.~Smith}}\ and\ \bibinfo {author} {\bibfnamefont {J.}~\bibnamefont {Gray}},\ }\bibfield  {title} {\bibinfo {title} {opt\_einsum - a python package for optimizing contraction order for einsum-like expressions},\ }\href {https://doi.org/10.21105/joss.00753} {\bibfield  {journal} {\bibinfo  {journal} {Journal of Open Source Software}\ }\textbf {\bibinfo {volume} {3}},\ \bibinfo {pages} {753} (\bibinfo {year} {2018})}\BibitemShut {NoStop}%
\bibitem [{\citenamefont {Boyd}\ and\ \citenamefont {Vandenberghe}(2004)}]{boyd2004convex}%
  \BibitemOpen
  \bibfield  {author} {\bibinfo {author} {\bibfnamefont {S.~P.}\ \bibnamefont {Boyd}}\ and\ \bibinfo {author} {\bibfnamefont {L.}~\bibnamefont {Vandenberghe}},\ }\href@noop {} {\emph {\bibinfo {title} {Convex optimization}}}\ (\bibinfo  {publisher} {Cambridge university press},\ \bibinfo {year} {2004})\BibitemShut {NoStop}%
\end{thebibliography}%

\clearpage
\newpage
\onecolumngrid
\appendix
\begin{center}
	\textbf{\large Supplementary Material for "Unified Framework for Calculating Convex Roof Resource Measures"}
\end{center}
\setcounter{equation}{0}
\setcounter{figure}{0}
\setcounter{table}{0}
\setcounter{page}{1}
\makeatletter
\renewcommand{\thetable}{S\arabic{table}}
\renewcommand{\thefigure}{S\arabic{figure}}
\renewcommand{\theequation}{S\arabic{equation}}

\newtheorem{theorem}{Theorem}
\newtheorem{definition}{Definition}

\section{Appendix A: Trivialization of the Stiefel manifold}

\begin{definition}[Trivialization \cite{lezcano2019trivializations}]
    A trivialization of a smooth manifold $\mathcal{M}$ is a surjective map from Euclidean space onto this manifold $g:\mathbb{R}^n\to \mathcal{M}$.
\end{definition}
To apply gradient-based optimization, the trivialization technique transforms optimization over the manifold $\mathcal{M}$ into that over Euclidean space, i.e., $\min_{x\in \mathcal{M}}f(x)=\min_{\theta\in \mathbb{R}^n}f(g(\theta))$. The  Stiefel manifold discussed in this work contains these complex matrices
\[ \mathrm{St}(n,r)=\left\{ X\in\mathbb{C}^{n\times r}:X^{\dagger}X=I_{r}\right\}, \]
satisfying their multiplication with their conjugate gives a $r$-by-$r$ identity matrix. The dimension of the Stiefel manifold $\mathrm{St}(n,r)$ is $2nr-r^2$. Below, we summarize three different trivialization maps for the Stiefel manifold.

\begin{theorem}[Matrix exponential as trivialization of $\mathrm{St}(n,r)$]
    The following map $g$ is a trivialization for the Stiefel manifold:
    \[ g(\theta)=\mathrm{column}_r\left(\mathrm{exp}\left(i\theta_0I_n+i\sum_{j=1}^{n^2-1}\theta_{j}M_{j}\right)\right):\mathbb{R}^{n^2}\to\mathrm{St}(n,r) \]
    where $\{M_i\}_{i=1}^{n^2-1}$ are $n$-by-$n$ generalized Gell-Mann matrices \cite{bertlmann2008bloch}, $I_n$ is a $n$-by-$n$ identity matrix and $\mathrm{column}_r(\cdot)$ denotes select the first $r$ columns of the matrix.
\end{theorem}

In the trivialization above, $n^2$ parameters in Euclidean space are used to construct unitary matrices, and then the first $r$ columns are selected as the Stiefel matrices. To prove its surjectivity, one should notice that any Stiefel matrix can be completed to a unitary matrix via concatenating the orthogonal space, thus validating it as a trivialization map. However, in practice, we often use $n^2-1$ parameters since the first parameter $\theta_0$ only brings a global phase, which will not affect the optimization. The back-propagation for the matrix exponential operation is already implemented in PyTorch framework \cite{pmlr-v97-lezcano-casado19a}.

\begin{theorem}[Polar projection as trivialization of $\mathrm{St}(n,r)$]
    The following map $g$, calculating the polar decomposition of the input matrix $A$, is a trivialization for Stiefel manifold:
    \[g(A)=A\left(A^{\dagger}A\right)^{-1/2}:\mathbb{C}^{n\times r}\to\mathrm{St}(n,r).\]
\end{theorem}

Above, the real and imaginary part of the domain $\mathbb{C}^{n\times r}$ can be viewed as Euclidean space separately, thus $2nr$ free parameters are needed for polar projection trivialization. For any Stiefel matrix $X\in\mathrm{St}(n,r)$, its pre-image contains at least itself $X\in g^{-1}(X)$, thus the surjectivity is guaranteed. The possible concern is that low-rank matrix $A$ with $\mathrm{rank}(A)<r$ is ill-defined for the polar projection map; However, these low-rank matrices are of measure-zero with respect to the domain $\mathbb{C}^{n\times r}$ and we find these matrices are never encountered during the optimization. In the forward evaluation, the matrix square root can be implemented via eigenvalue decomposition:
\[ X=U\Sigma U^{\dagger}\;\Rightarrow\;X^{1/2}=U\Sigma^{1/2}U^{\dagger}, \]
with eigen vector matrix $U$ and diagonal eigenvalue $\Sigma$. To avoid the degeneracy issue of the eigenvalue decomposition, the gradient back-propagation for the matrix square root is calculated via solving a Sylvester equation \cite{cardoso2012computation}, which we used the implementation from SciPy package \cite{2020SciPy-NMeth}.

\begin{theorem}[Euler-Hurwitz angles as trivialization of $\mathrm{St}(n,r)$ \cite{rothlisberger2009numerical}]
    The following map $g:\mathbb{R}^{m_{r}+n-r}\times\mathbb{R}^{m_{r}+n-r}\times\mathbb{R}^{r}\to\mathrm{St}(n, r)$ is a trivialization for the Stiefel manifold
    \begin{equation}\label{eq:stiefel-euler-param}
        g(\theta,\phi,\varphi)=\left(\prod_{i=1}^{n-1}G^{(n-i)}\left(\theta_{i},\phi_{i}\right)\right)^{\dagger}\cdots\left(\prod_{i=m_{r}+1}^{m_{r}+n-r}G^{(m_{r}+n-i)}\left(\theta_{i},\phi_{i}\right)\right)^{\dagger}D(\varphi),
    \end{equation}
    \begin{equation}
        D(\varphi)=\mathrm{diag}\left\{ e^{i\varphi_{1}},\cdots,e^{i\varphi_{r}}\right\} :\mathbb{R}^{r}\to\mathbb{C}^{n\times r},
    \end{equation}
    with $m_{r}=nr-n-r^{2}/2+r/2$ and the Givens rotation matrix $G_{i,j}^{(s)}(\theta,\phi)\in\mathbb{C}^{n\times n}$ as:
    \[ G_{i,j}^{(s)}(\theta,\phi)=\begin{cases}
e^{i\phi}\cos\theta, & \mathrm{if}~i=j=s\\
e^{-i\phi}\sin\theta, & \mathrm{if}~i=s,j=s+1\\
-e^{i\phi}\sin\theta,& \mathrm{if}~i=s+1,j=s\\
e^{-i\phi}\cos\theta, & \mathrm{if}~i=j=s+1\\
\delta_{ij}, & \mathrm{otherwise}.
\end{cases} \]
\end{theorem}
To prove the mapping above is a valid trivialization, we must verify that every Stiefel matrix $X\in\mathrm{St}(n,r)$ has at least one pre-image. We first multiply $X$ with a Givens matrix from the left-hand side
\[ \tilde{X}^{(1)}=G^{(n-1)}\left(\theta_{1},\phi_{1}\right)X, \]
\[ \theta_{1}=\arctan\left|\frac{X_{n,1}}{X_{n-1,1}}\right|,\phi_{1}=\frac{1}{2}\arg\left(\frac{X_{n,1}}{X_{n-1,1}}\right), \]
to eliminate the element $\tilde{X}_{n,1}^{(1)}=0$. When $X_{n-1,1}=0$, the parameters can be chosen as $\left(\theta_{1},\phi_{1}\right)=\left(\pi/2,0\right)$. For simplicity, we assume the denominators below are nonzero, otherwise, the same convention as here can be used. As Givens matrix is unitary, the resulting matrix $\tilde{X}^{(1)}$ is also a Stiefel matrix. To eliminate one more element, another Givens matrix is required as below:
\[ \tilde{X}^{(2)}=G^{(n-2)}\left(\theta_{2},\phi_{2}\right)G^{(n-1)}\left(\theta_{1},\phi_{1}\right)X, \]
\[ \theta_{2}=\arctan\left|\frac{\tilde{X}_{n-1,1}^{(1)}}{\tilde{X}_{n-2,1}^{(1)}}\right|,\phi_{1}=\frac{1}{2}\arg\left(\frac{\tilde{X}_{n-1,1}^{(1)}}{\tilde{X}_{n-2,1}^{(1)}}\right). \]
For this new Stiefel matrix, two elements are zero $\tilde{X}_{n,1}^{(2)}=\tilde{X}_{n-1,1}^{(2)}=0$. We can continue the process of removing almost all elements in the first column:
\[ \tilde{X}^{(n-1)}=\left(\prod_{i=1}^{n-1}G^{(n-i)}\left(\theta_{i},\phi_{i}\right)\right)X. \]
The first column of $\tilde{X}^{(n-1)}$ is
\[ \left|\tilde{X}_{1,1}^{(n-1)}\right|=1,\tilde{X}_{2,1}^{(n-1)}=\tilde{X}_{3,1}^{(n-1)}=\cdots=\tilde{X}_{n,1}^{(n-1)}=0, \]
as it's a unit-length vector. Similarly, we can eliminate most entries in other columns
\begin{equation}
    \tilde{X}^{\left(m_{r}+n-r\right)}=\left(\prod_{i=m_{r}+1}^{m_{r}+n-r}G^{(m_{r}+n-i)}\left(\theta_{i},\phi_{i}\right)\right)\cdots\left(\prod_{i=1}^{n-1}G^{(n-i)}\left(\theta_{i},\phi_{i}\right)\right)X,
\end{equation}
which is a diagonal matrix $\tilde{X}_{i,j}^{\left(m_{r}+n-r\right)}=\delta_{ij}e^{i\varphi_{i}}$. Sending all Givens matrices to the left-hand side, we will arrive at equation (\ref{eq:stiefel-euler-param}), thus finding one pre-image of Stiefel matrix $X$.

In conclusion, the numbers of parameters required for various trivializations onto the Stiefel manifold $\mathrm{St}(n,r)$ are different. For the matrix exponential, $n^2-1$ parameters are necessary. The polar projection method requires $2nr$ parameters, while the Euler-Hurwitz angles method needs $2nr - r^2$ parameters. It is important to note that the number of parameters does not directly dictate the time efficiency of a particular trivialization in optimization processes.

\section{Appendix B: Optimization of the convex roof resource measure}
Convex roof resource measure begins with a measure of pure states and then extends to the mixed states as follows:
\begin{equation}
    R(\rho)=\min_{\{p_i,|\psi_i\rangle\}}\sum_i p_i R(|\psi_i\rangle),
\end{equation}
where the minimization is taken over all possible pure-state decompositions of the given mixed state $\rho$ satisfying $\rho =\sum_i p_i |\psi_i\rangle \langle \psi_i|$. In the main text, we conclude that for a rank-$r$ state, the optimization problem related to $n$-entry pure-state decomposition can be transferred to the optimization over the Stiefel manifold $\mathrm{St}(n,r)$, i.e.,
\[
    \min_{\{p_i,|\psi_i\rangle\}} \sum_{i=1}^n p_i R(|\psi_i\rangle)= \min_{X \in \operatorname{St}(n,r)} \sum_{i=1}^n p_i(X)R(|\psi_i(X)\rangle).
\]
This transformation can be realized by introducing a series of unnormalized auxiliary states $\{ |\tilde{\psi}_i\rangle \}$ satisfying $p_i= \langle \tilde{\psi}_i|\tilde{\psi}_i\rangle$ and $|\psi_i\rangle = \frac{1}{\sqrt{p_i}}|\tilde{\psi}_i\rangle$. These auxiliary states can be represented by Stiefel matrices as follows:
\begin{equation}
    |\tilde{\psi}_i\rangle = \sum_{j=1}^r \sqrt{\lambda_j} X_{ij} |\lambda_j\rangle,
\label{eq:auxiliary}
\end{equation}
with $\left\{\lambda_i,|\lambda_i\rangle\right\}$ being the eigenvalue and eigenvector of the input density matrix $\rho$.

For the consistency, we choose the same hyper-parameters unless otherwise stated, where the number of pure states for the decomposition is $n=2d$ ($d$ is the dimension of the given Hilbert), the repeat number for each optimization is $N=3$ for alleviating the local minimum issue and the tolerance of the optimization is set to be $10^{-14}$. And for convenience, in practice, we use $\mathrm{St}(n,d)$ instead of $\mathrm{St}(n,r)$ since the eigenvalue decomposition gives us a whole series of eigenvectors with corresponding eigenvalues which can be zero. All the results are produced on a standard laptop computer.

\subsection{Entanglement of formation}
One of the most famous measures, the entanglement of formation (EoF) is given by  $E_f(|\psi\rangle)=S(\operatorname{Tr}_B[|\psi\rangle\langle\psi|])$ for bipartite pure state $|\psi\rangle\in\mathcal{H}_A\otimes \mathcal{H}_B$, where $\operatorname{Tr}_B$ denotes the partial trace over the subsystem B and $S(\rho)=-\operatorname{Tr}[\rho \ln \rho]$ is the Von Neumann entropy of a density matrix $\rho$. For mixed state $\rho$, EoF is defined via convex roof extension $E_f(\rho)=\min_{\{p_i,|\psi_i\rangle\}}\sum_i p_i E_f(|\psi_i\rangle)$ satisfying ensemble decomposition $\rho=\sum_i p_i|\psi_i\rangle\langle \psi_i|$. To address potential numerical instability, the explicit objective function is written down as below:
\begin{equation}\label{eq:eof-objective}
   \mathcal{L}=\sum_i p_i E_f(|\psi_i\rangle)=-\sum_ip_i\mathrm{Tr}[\rho_i\ln\rho_i]=\sum_{i}p_{i}\ln p_{i}-\mathrm{Tr}[\tilde{\rho}_{i}\ln\tilde{\rho}_{i}].
\end{equation}
Above, the objective function is finally expressed in the probability $p_{i}=\langle\tilde{\psi}_{i}|\tilde{\psi}_{i}\rangle$ and unnormalized reduced density matrix $\tilde{\rho}_{i}=\mathrm{Tr}_{B}\left[|\tilde{\psi}_{i}\rangle\langle\tilde{\psi}_{i}|\right]$, with the auxiliary states $\{ |\tilde{\psi}_i\rangle \}$ shown in equation (\ref{eq:auxiliary}).
The normalized quantity $\rho_{i}=\mathrm{Tr}_{B}\left[|\psi_{i}\rangle\langle\psi_{i}|\right]$ are less preferred since the normalization operation, for example, $|\psi_{i}\rangle=\frac{1}{\sqrt{p_{i}}}|\tilde{\psi}_{i}\rangle$, might lead to zero-division numerical instability. 

Another operation that could cause numerical instability is the logarithm in equation (\ref{eq:eof-objective}). The matrix logarithm is implemented via eigenvalue decomposition $\operatorname{Tr}[\tilde{\rho}_i \ln \tilde{\rho}_i]=\sum_{j}\tilde{\lambda}_{ij}\ln \tilde{\lambda}_{ij}$ with $\{\tilde{\lambda}_{ij}\}$ being the eigenvalue of $\tilde{\rho}_i$. In this way, both two terms in the equation (\ref{eq:eof-objective}) are like the form $x\ln x$. Although the limit value $\lim_{x\to 0^+} x\ln x=0$ is a finite, numerical evaluation of $\ln(x)$ for some almost zero value $x\to 0$ will lead to the numerical error: usually, $\ln(0)$ gives the value "inf", then multiplying "0" with "inf" gives Not-A-Number (NAN) error which terminates the whole optimization. Even worse for the gradient-based optimization, the gradient of $x\ln(x)$ diverges when $x$ approaches zero. To avoid these issues, we implemented $x\ln (x)$ in a truncated way as below:
\begin{equation}\label{eq:xlnx-approximation}
    x\ln x\approx\begin{cases}
x\ln x, & x>\varepsilon\\
x\ln(\varepsilon), & x\leq\varepsilon
\end{cases},
\end{equation}
When $x$ is smaller than some small value $\varepsilon$, the function value is set to $x\ln(\varepsilon)$. We choose $\varepsilon$ as the smallest positive double-precision floating-point value in our numerical experiments.

\begin{table}[h]
\centering
\caption{\label{table:eof}The average computational time (in second) for computing EoF of $d_A \otimes d_B$ Haar-random states with rank $r$, using different trivialization mappings. Each average computational time is obtained from 10 randomly generated states.}
\begin{tabular}{c|c|c|c|c}
    \toprule
    $\left(d_A, d_B\right)$ & $r$ & Polar projection & Matrix exponential & Euler-Hurwitz angles \\
    \midrule \multirow{4}{*}{$(2,2)$} & 1 & 0.003 & 0.005 & 0.021 \\
    \cmidrule { 2 - 5 } & 2 & 0.050 & 0.094 & 1.453 \\
    \cmidrule { 2 - 5 } & 3 & 0.106 & 0.225 & 4.131 \\
    \cmidrule { 2 - 5 } & 4 & 0.160 & 0.388 & 8.461 \\
    \midrule \multirow{6}{*}{$(2,3)$} & 1 & 0.003 & 0.006 & 0.045 \\
    \cmidrule { 2 - 5 } & 2 & 0.051 & 0.096 & 22.994 \\
    \cmidrule { 2 - 5 } & 3 & 0.179 & 0.422 & 47.991 \\
    \cmidrule { 2 - 5 } & 4 & 0.429 & 1.255 & 30.211 \\
    \cmidrule { 2 - 5 } & 5 & 3.407 & 7.495 & 79.208 \\
    \cmidrule { 2 - 5 } & 6 & 7.138 & 22.795 & 325.547 \\
    \bottomrule
\end{tabular}
\end{table}

Above, we compare different trivializations for calculating the Entanglement of Formation (EoF), as shown in Table~\ref{table:eof}. We observe that the polar projection demonstrates superior performance in terms of time efficiency compared to other approaches. This phenomenon is consistently observed across various scenarios in practice, leading us to select the polar projection as our primary trivialization mapping.

\subsection{Linear entropy of entanglement}

Linear entropy of entanglement is another important entanglement measure. To calculate linear entropy of entanglement via optimization, the involved objective function $\mathcal{L}$ is
\begin{equation}\label{eq:linear-ent-obj}
    \mathcal{L}=\sum_{i}p_{i}E_{l}\left(|\psi_{i}\rangle\right)=1-\sum_{i}p_{i}\mathrm{Tr}\left[\rho_{i}^{2}\right]=1-\sum_{i}\frac{1}{p_{i}}\mathrm{Tr}\left[\tilde{\rho}_{i}^{2}\right].
\end{equation}
As in equation (\ref{eq:eof-objective}), we express the formula in terms of the probability $p_i=\langle \tilde{\psi}_i|\tilde{\psi}_i\rangle$, and unnormalized reduced density matrix $\tilde{\rho}_{i}=\mathrm{Tr}_{B}\left[|\tilde{\psi}_{i}\rangle\langle\tilde{\psi}_{i}|\right]$ with the auxiliary states $\{ |\tilde{\psi}_i\rangle \}$ shown in equation (\ref{eq:auxiliary}).  However, the zero-division instability could appear for this new expression: as $|\tilde{\psi}_i\rangle$ approaches zero, both nominator $p_i$ and denominator $\mathrm{Tr}[\tilde{\rho}_i^2]$ go to zero, and their quotient will cause numerical instability. Since the denominator goes to zero faster than the nominator part, we set the denominator to $\varepsilon$ numerically when denominator is smaller than a threshold $\varepsilon$, which is similar to the equation (\ref{eq:xlnx-approximation}). Also, the evaluation of the objective function $\mathcal{L}$ can be viewed as the contraction of the relevant tensors, so that we use package "opt\_einsum" for easier implementation and better performance \cite{Smith_opt_einsum}.

In the previous study \cite{toth2015evaluating}, a powerful method was proposed to compute the lower bounds of the linear entropy of entanglement base on semi-definite programming (SDP). It reformulates the computation of linear entropy of entanglement $E_l(\rho)$ as follows:
\begin{equation}\label{eq:ppt}
    \begin{aligned}
        \min_{\omega_{ABA'B'}}&\quad\mathrm{Tr}\left[\mathcal{A}_{AA'}\otimes\mathbb{I}_{BB'}\omega_{ABA'B'}\right]\\
        \mathrm{s.t.}&\begin{cases}
            \omega_{ABA'B'}\succeq0\\
            \mathrm{Tr}_{A'B'}\left[\omega_{ABA'B'}\right]=\rho\\
            \omega_{ABA'B'}\text{ is bosonic and separable for AB and A'B'}
        \end{cases}
        \end{aligned}
\end{equation}
The projector to the anti-symmetric space is defined as $\mathcal{A}_{\mathrm{AA}^{\prime}}:=(\mathbb{I}-\mathcal{F})_{\mathrm{AA}^{\prime}}$, where $\mathcal{F}$ is the flip operator. Then the separability constraint can be further relaxed to the positive partial transpose (PPT). 

\begin{table}[h]
\centering
\caption{\label{table:lee}The average computational time (in seconds) for computing the linear entropy of entanglement of $d_A \otimes d_B$ Haar-random states with rank $r$, using polar projection and PPT relaxation. Each average computational time is obtained from 10 randomly generated states. NA means "not available in an acceptable time".}
\begin{tabular}{|c|c|ccc|cccc|cccc|}
    \toprule
    $(d_A,d_B)$    & $(2,2)$ & \multicolumn{3}{c|}{$(2,3)$}                                      & \multicolumn{4}{c|}{$(3,3)$}                                                                     & \multicolumn{4}{c|}{$(3,4)$}                                                                   \\ \hline
    Rank $r$       & $4$     & $2$     & $4$     & $6$     & $1$     & $3$      & $6$      & $9$     & $3$     & $6$     & $9$     & $12$    \\ \hline
    Polar projection & $0.134$ & $0.048$ & $0.605$ & $8.380$ & $0.005$ & $0.122$  & $0.529$  & $7.976$ & $0.169$ & $0.344$ & $1.259$ & $5.963$ \\ \hline
    PPT relaxation   & $0.114$ & $0.538$ & $0.640$ & $0.604$ & $6.396$ & $13.776$ & $65.825$ & $9.758$ & NA    & NA    & NA    & NA    \\ \bottomrule
\end{tabular}
\end{table}
In the Table~\ref{table:lee}, we compare the time efficiency of the polar projection and PPT relaxation methods for computing the linear entropy of entanglement. From equation (\ref{eq:ppt}), we observe that the size of the variable $\omega$ is $(d_A^2 d_B^2) \times (d_A^2 d_B^2)$, which entails a requirement of $O(d_A^4 d_B^4)$ parameters. This substantial memory demand significantly restricts its application to larger dimensions, which is consistent with the results we obtained. Conversely, gradient-based optimization only requires $4d_A^2 d_B^2$ parameters.

\subsection{Geometric measure of coherence}

The objective function for calculating the geometric measure of coherence is given by:
\begin{equation}\label{eq:obj-g-coherence}
    \mathcal{L} = 1 - \sum_{i} \max_{j \in \{0, 1, \ldots, d-1\}} \tilde{\rho}_{i,jj},
\end{equation}
where $\tilde{\rho}_{i,jj}$ is the $j$-th diagonal component of the $i$-th entry of the ensemble decomposition, defined as $\tilde{\rho}_{i,jj} = \langle j | \tilde{\psi}_{i} \rangle \langle \tilde{\psi}_{i} | j \rangle$. The unnormalized auxiliary state $| \tilde{\psi}_{i} \rangle$ is the same as in equation (\ref{eq:auxiliary}).

The objective function involves a maximum operation over the index $j$, which can lead to a discontinuous gradient when two entries have the same maximum value, i.e., $\max_j \tilde{\rho}_{i,jj} = \tilde{\rho}_{i,\mu\mu} = \tilde{\rho}_{i,\nu\nu}$ where $\mu \neq \nu$. This discontinuity can hinder gradient-based optimization methods, especially the L-BFGS optimizer.

To address this issue, we approximate the maximum operation using the log-sum-exp operation \cite{boyd2004convex}:
\[ 
\max_{j \in \{0, 1, \ldots, d-1\}} \tilde{\rho}_{i,jj} = \lim_{T \to 0^{+}} T \log\left(\sum_{j=0}^{d-1} \exp\left(\tilde{\rho}_{i,jj} / T\right)\right).
\]
As the temperature \(T\) approaches zero, the log-sum-exp operation converges to the exact maximum. In practice, we perform the optimization with an empirical temperature, then evaluate the objective function using the maximum operation on the optimized parameters, as reported in the main text. We used a hyper-parameter value of \(T = 0.3\).

\begin{figure}
    \centering
    \includegraphics[width=0.6\textwidth]{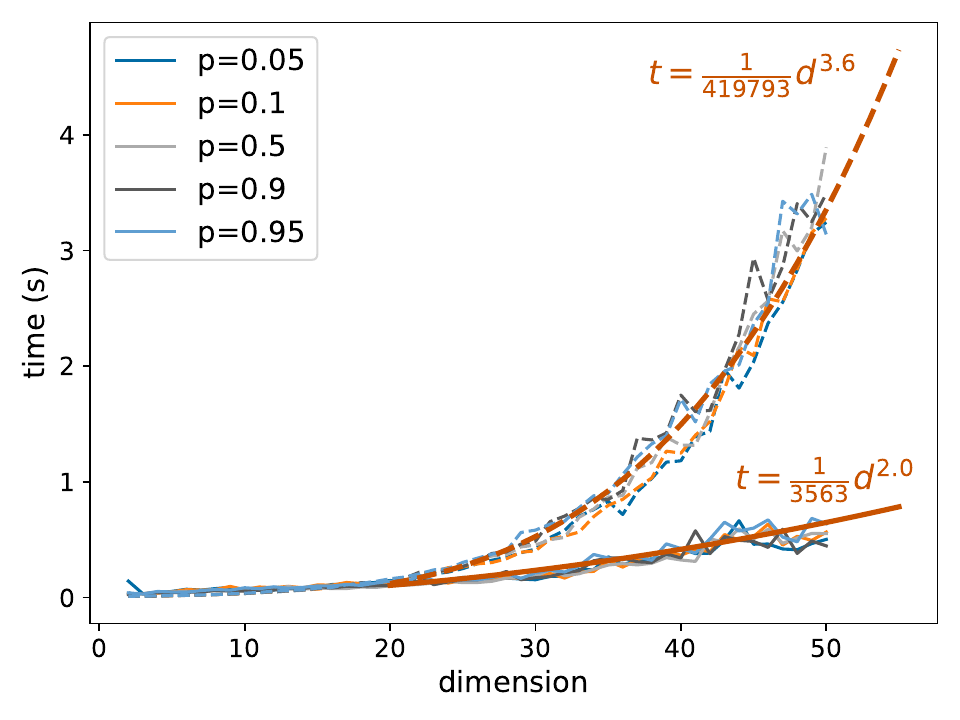}
    \caption{Computational times (in seconds) for calculating \(C_g\) of noisy coherent states indexed by the purity \(p\) with different dimensions \(d\). The solid lines represent the results of the gradient-based optimization, while the dashed lines represent the results of the SDP method. The final fitting function \(t = \alpha d^{\beta}\), with \(\beta = 2\) for the gradient-based optimization and \(\beta = 3.6\) for the SDP method, is denoted by the red lines.}
    \label{fig:gmc_time}
\end{figure}

In \cite{zhang2020numerical}, the authors proposed a method to calculate the geometric measure of coherence $C_g(\rho)$ for a given state $\rho$ in the Hilbert space $\mathcal{H}$, using the semi-definite programming (SDP) method, which is formulated as:

\begin{equation}\label{eq:gmc_sdp}
    \begin{aligned}
        \max_{X} & \quad \frac{1}{2} \operatorname{Tr}[\mathrm{X}]+\frac{1}{2} \operatorname{Tr}[\mathrm{X}^{\dagger}]\\
        \mathrm{s.t.}&\begin{cases}
            \left(\begin{array}{cc}
                \rho & X \\
                X^{\dagger} & \chi
            \end{array}\right) \succeq 0, \\
            X \in L(\mathcal{H}), \\
            \operatorname{Tr}[\chi] = 1,\chi \succeq 0, \chi~ \text{is diagonal},
            \end{cases}
    \end{aligned}
\end{equation}
where $L(\mathcal{H})$ denotes the collection of all linear mappings in the Hilbert space $\mathcal{H}$. In the main text, we calculate the geometric measure of coherence for noisy coherent states across different dimensions $d$. Here, we analyze the time-scaling behavior of the gradient-based optimization and the SDP method for these computations, as illustrated in Fig.~\ref{fig:gmc_time}. We fit the data from dimensions \(d = 20\) to \(d = 50\) using the function \(t = \alpha d^{\beta}\), where the exponent \(\beta = 2\) for the gradient-based optimization and \(\beta = 3.6\) for the SDP method. The gradient-based optimization demonstrates superior time-scaling behavior compared to the SDP method, particularly for larger dimensions.

\subsection{Stabilizer purity}

In this subsection, we focus on stabilizer purity, denoted as $P_\alpha(\rho)$. Other measures of magic-state resources, such as the $\alpha$-R\'{e}nyi stabilizer entropy $M_\alpha(\rho)$ and the linear stabilizer entropy $M^{\mathrm{lin}}_\alpha(\rho)$, can be directly calculated from the stabilizer purity. 

The objective function to determine the stabilizer purity for an $n$-qubit system is given by:
\begin{equation}
    \mathcal{L} = -2^{-n} \sum_{i} p_{i}^{1-2\alpha} \sum_{P \in \mathbb{P}_{n}} \left| \langle \tilde{\psi}_{i} | P | \tilde{\psi}_{i} \rangle \right|^{2\alpha}.
\end{equation}
Here, the negative sign is included because stabilizer purity for a mixed state is obtained by maximizing over all pure-state decompositions, while the objective function $\mathcal{L}$ needs to be minimized.

The unnormalized auxiliary state $|\tilde{\psi}_i\rangle$ is defined in equation (\ref{eq:auxiliary}). Similar to the linear entropy of entanglement described in equation (\ref{eq:linear-ent-obj}), this formulation also encounters zero-division issues when $|\tilde{\psi}_i\rangle$ becomes the zero vector. To avoid numerical instability, we manually set the denominator to $\varepsilon$ if $p_i^{2\alpha-1}$ is smaller than a threshold $\varepsilon$, as described in equation (\ref{eq:xlnx-approximation}).

\end{document}